\documentclass{article_saj}
\usepackage[utf8]{inputenc}
\usepackage{natbib}
\usepackage{graphicx,saj,multicol,subeqnarray}
\usepackage{subcaption}
\usepackage{float}
\usepackage{xcolor}
\usepackage{widetext}
\usepackage{url}
\usepackage{bm}
\usepackage{placeins}
\usepackage{tikz} 
\usepackage{pifont} 
\usepackage{amsfonts}
\usepackage{amssymb}
\usepackage{amsmath,upgreek}
\usepackage{titlesec}

\def\point#1{\hbox{\setbox7=\hbox to0.6em{\hfil.\hfil}%
\setbox8=\hbox to0.5em{\hfil$^{#1}$\hfil}%
\box7\kern-0.5em\box8}}

\titlelabel{\thetitle.\quad}
\definecolor{xlinkcolor}{cmyk}{1,0.6,0,0}
\usepackage{amsmath}
\usepackage[bookmarks=false,         
     pdfnewwindow=true,      
     colorlinks=true,    
     linkcolor=xlinkcolor,     
     citecolor=xlinkcolor,     
     filecolor=xlinkcolor,  
     urlcolor=xlinkcolor,      
final=true
]{hyperref}
\usepackage{xspace}
\usepackage[nolist]{acronym}
\usepackage{multicol}
\usepackage{supertabular}
\usepackage{multirow}
\usepackage{comment}

\newcommand{\ergcm}[1]{erg~cm$^{-2}$ s$^{-1}$}

\newcommand{\HII}{H{\sc ii}}

\def\arcsec{\hbox{$^{\prime\prime}$}}

\newcommand{\farcs}{\mbox{\ensuremath{.\!\!^{\prime\prime}}}}

\begin{acronym}[AWGN]

\acro{2MASS}{Two Micron All Sky Survey}
\acro{2MASX}{Two Micron All Sky Survey Extended Source Catalogue}
\acro{30Dor}[30 Dor]{30~Doradus}
\acro{AeReS}{Aegean Residual}
\acro{AGN}{active galactic nuclei}
\acro{ARC}{Australian Research Council}
\acro{ATCA}{Australia Telescope Compact Array}
\acro{ATESP}{Australia Telescope ESO Slice Project}
\acro{ATNF}{Australia Telescope National Facility}
\acro{ATOA}{Australia Telescope Online Archive}
\acro{AT20G}{Australia Telescope 20 GHz Survey}
\acro{ASKAP}{Australian Square Kilometre Array Pathfinder}
\acro{BCE}{bright compact extended}
\acro{BETA}{Boolardy Engineering Test Array}
\acro{BGG}{Brightest Galaxy Group}
\acro{BL}[BL Lac]{BL Lacertae Objects}
\acro{CASS}{CSIRO Astronomy and Space Science}
\acro{CABB}{Compact Array Broadband Back-end}
\acro{CC}{core-collapse}
\acro{CHII}[{\sc CHii}]{Compact \textsc{Hii}}
\acro{CIE}{collisional ionisation equilibrium}
\acro{CMB}{Cosmic Microwave Background}
\acro{CSIRO}{Australian Commonwealth Scientific and Industrial Research Organisation}
\acro{CSS}{Compact Steep Spectrum}
\acro{DESI}{Dark Energy Spectroscopic Instrument}
\acro{DS9}[\textsc{DS9}]{\textsc{SAOImage DS9}}
\acro{DSS}{Digital Sky Survey}
\acro{DSA}{diffuse shock acceleration}
\acro{EBHIS}{Effelsberg--Bonn H\,{\sc i} Survey}
\acro{EM}{Electromagnetic}
\acro{EMU}{Evolutionary Map of the Universe}
\acro{ev}[eV]{electronvolt\acroextra{: 1 eV $\approx 1.6 \times 10^{-19}$ J}}
\acro{eROSITA}{extended R\"Ontgen Survey with an Imaging Telescope Array}
\acro{Fermi-LAT}{Fermi Large Area Telescope}
\acro{FIR}{Far Infrared}
\acro{FITS}[\textsc{Fits}]{Flexible Image Transport System}
\acro{FRB}{Fast Radio Bursts}
\acro{FSRQ}{Flat Spectrum Radio Quasars}
\acro{FWHM}{Full Width at Half-Maximum}
\acro{GASS}{Galactic All-Sky Survey}
\acro{GLEAM}{GaLactic and Extragalactic All-sky MWA}
\acro{GPS}{Gigahertz Peak Spectrum}
\acro{HEMT}{High Electron Mobility Transistor}
\acro{HERITAGE}{HERschel Inventory of The Agents of Galaxy Evolution}
\acro{H.E.S.S.}{High Energy Stereoscopic System}
\acro{HFP}[HFP]{High-Frequency Peaker}
\acro{HPBW}{Half Power Beam Width} 
\acro{HzRGs}{High Redshift Radio Galaxies}
\acro{pHFP}[pHFP]{Potential High Frequency Peaker}
\acro{HI}[H{\sc i}]{Neutral Atomic Hydrogen} 
\acro{HST}{\textit{Hubble Space Telescope}}
\acro{ICM}{intracluster medium}
\acro{IAU}{International Astronomical Union}
\acro{IFRSs}{Infrared Faint Radio Sources}
\acro{ISM}{interstellar medium}
\acro{IFRS}{Infrared Faint Radio Source}
\acro{IR}{Infrared}
\acro{JY}[Jy]{Jansky\acroextra{, 1 Jy = $10^{-26} \times \mathrm{W~ m}^{-2}~\mathrm{Hz}^{-1}$}} 
\acro{LLS}{Largest Linear Size}
\acro{LFAA}{Low-Frequency Aperture Array}
\acro{LMC}{Large Magellanic Cloud}
\acro{LSO}[LSOs]{Large Scale Objects}
\acro{MACHO}{Massive Astrophysical Compact Halo Objects}
\acro{MAGMA}{Magellanic Mopra Assessment}
\acro{MC}[MC]{Magellanic Cloud}
\acro{MS}[MS]{Magellanic Stream} 
\acro{MB}[MB]{Magellanic Bridge} 
\acro{MCELS}{Magellanic Cloud Emission Line Survey}
\acro{MW}{Milky Way}
\acro{MIPS}{Multiband Imaging Photometer}
\acro{MIRIAD}[\textsc{Miriad}]{Multichannel Image Reconstruction, Image Analysis and Display}
\acro{MIT}{Massachusetts Institute of Technology}
\acro{MM}{mixed-morphology}
\acro{MOST}{Molonglo Observatory Synthesis Telescope}
\acro{MQS}{Magellanic Quasars Survey}
\acro{MRC}{Molonglo Reference Catalogue of Radio Sources}
\acro{MWA}{Murchison Widefield Array}
\acro{NED}{NASA/IPAC Extragalactic Database}
\acro{NRAO}{National Radio Astronomy Observatory}
\acro{NVSS}{NRAO VLA Sky Survey}
\acro{OPAL}{Online Proposal Applications \& Links}
\acrodefplural{ORC}{Odd Radio Circles}
\acro{ORC}{Odd Radio Circle}
\acro{OVV}{Optically Violent Variable Quasars}
\acro{PACS}{Photodetector Array Camera and Spectrometer}
\acro{PAF}{Phased Array Feed}
\acro{pc}{parsec\acroextra{: 1 pc $\simeq 3.09 \times 10^{16}$ m}}
\acro{PMN}{Parkes-MIT-NRAO}
\acro{PN}{planetary nebula}
\acrodefplural{PN}[PNe]{planetary nebulae}
\acro{PWN}{pulsar wind nebula}
\acrodefplural{PWN}[PWNe]{pulsar wind nebulae}
\acro{RN}{reflection nebula}
\acrodefplural{RN}[RNe]{reflection nebulae}
\acro{QSO}{Quasi-Stellar Object}
\acro{RA}{Right Ascension}
\acro{RFI}{Radio-Frequency Interference}
\acro{RMS}[rms]{Root Mean Squared}
\acro{SAGE}{Surveying the Agents of a Galaxy’s Evolution}
\acro{SARAO}{South African Radio Astronomy Observatory}
\acro{SDSS}{Sloan Digital Sky Survey}
\acro{SED}{spectral energy distribution}
\acro{SI}[$\alpha$]{Spectral Index\acroextra{, $S \propto \nu^\alpha$}}
\acro{SHS}{SuperCOSMOS H$\alpha$ Survey}
\acro{SKA}{Square Kilometre Array}
\acro{SMB}{Super Massive Blackholes}
\acro{SMC}{Small Magellanic Cloud}
\acro{SN}{Supernova}
\acro{SPIRE}{Spectral and Photometric Imaging Receiver}
\acro{SUMSS}{Sydney University Molonglo Sky Survey}
\acro{TOPCAT}[\textsc{Topcat}]{Tool for OPerations on Catalogues And Tables}
\acro{USS}{Ultra Steep Spectrum}
\acro{YSOs}{young stellar objects}
\acro{WBAC}{Wide-Band Analogue Correlator}
\acro{WIFES}[WiFeS]{Wide-Field Spectrograph}
\acro{WISE}{Wide-Field Infrared Survey Explorer}
\acro{VLBI}{Very Long Baseline Interferometry} 
\acro{VLSR}[\textbf{$v_{lsr}$}]{Velocity in the Line of Sight}
\acro{SMASH}{Survey of the Magellanic Stellar History}
\acro{SNR}{supernova remnant}
\acro{SD}{single-degenerate}
\acro{SUMSS}{Sydney University Molonglo Sky Survey}
\acro{SMBH}{Super Massive Black Hole}
\acro{CARTA}{Cube Analysis and Rendering Tool for Astronomy}
\acro{NEI}{non-equilibrium ionisation}
\acro{UV}{ultra-violet}
\acro{FUSE}{Far Ultraviolet Spectroscopic Explorer}
\acro{RM}{rotation measure}
\acro{H.E.S.S.}{High Energy Spectroscopic System}
\acro{CTA}{Cherenkov Telescope Array}
\acro{POSSUM}{POlarisation Sky Survey of the Universe's Magnetism}
\acro{ASKAP}{Australian Square Kilometre Array Pathfinder}
\acro{BANE}{Background and Noise Estimator}
\acro{MWA}{Murchison Widefield Array}
\acro{POSSUM}{POlarisation Sky Survey of the Universe's Magnetism}
\acro{RMS}{Root Mean Squared}
\acro{SMC}{Small Magellanic Cloud}
\acro{MeerKAT}{Meer-Karoo Array Telescope}
\acro{ISM}{interstellar medium}
\acro{Milliquas}{The Million Quasars}
\acro{DESI}{Dark Energy Spectroscopic Instrument}
\acro{PN}{planetary nebula}
\acro{YSO}{young stellar object}
\acro{QSO}{quasi-stellar object}
\acro{AGN}{active galactic nuclei}
\acro{CSIRO}{Commonwealth Scientific and Industrial Research Organisation}
\acro{G}{galaxy}
\acro{Rad}{radio source}
\acro{rG}{radio galaxy}
\acro{RB?}{radio binary}
\acro{X}{X-ray source}
\acro{Psr}{pulsar}
\acro{Em*}{emission-line galaxy}
\acro{SyG}{Seyfert galaxy}
\acro{Cl*}{cluster of stars}
\acro{WD}{white dwarf}
\acrodefplural{WD}[WDs]{white dwarfs}

\end{acronym}

\def\papertype{\ \hfill\ Editorial}

\def\udc{52}
\begin{document}
\parindent=.5cm
\baselineskip=3.8truemm
\columnsep=.5truecm
\newenvironment{lefteqnarray}{\arraycolsep=0pt\begin{eqnarray}}
{\end{eqnarray}\protect\aftergroup\ignorespaces}
\newenvironment{lefteqnarray*}{\arraycolsep=0pt\begin{eqnarray*}}
{\end{eqnarray*}\protect\aftergroup\ignorespaces}
\newenvironment{leftsubeqnarray}{\arraycolsep=0pt\begin{subeqnarray}}
{\end{subeqnarray}\protect\aftergroup\ignorespaces}
%


\markboth{\eightrm Radio-Continuum Survey of \ac{SMC}} 
{\eightrm O. K. KHATTAB {\lowercase{\eightit{et al.}}}}

\begin{strip}

{\ }

\vskip-1cm

\publ

\type

{\ }


\title{New ASKAP radio-continuum surveys of the Small Magellanic Cloud}


\authors{
O. K. Khattab$^1$,
M. D. Filipovi\'c$^1$,
Z. J. Smeaton$^1$,
R. Z. E. Alsaberi$^{2,1}$,
E. J. Crawford$^{1}$,
}
\authors{
D. Leahy$^3$,
S. Dai$^{4,1}$
and N. Rajabpour$^1$
}

\vskip3mm


\address{$^1$Western Sydney University, Locked Bag 1797, Penrith South DC, NSW 2751, Australia}
\address{$^2$Faculty of Engineering, Gifu University, 1-1 Yanagido, Gifu 501-1193, Japan}
\address{$^3$Department of Physics and Astronomy, University of Calgary, Calgary, Alberta, T2N IN4, Canada}
\address{$^4$Australian Telescope National Facility, CSIRO, Space and Astronomy, P.O. Box 76, Epping, NSW 1710, Australia}

\Email{22197951@student.westernsydney.edu.au}


\dates{XXX}{XXX}


\abstract{We present two new radio-continuum images from the \ac{ASKAP} \ac{POSSUM} survey in the direction of the \ac{SMC}. The two new source lists produced from these images contain 36,571 radio continuum sources observed at 944\,MHz and 15,227 sources at 1367\,MHz, with beam sizes of 14\farcs5$\times$12\farcs2 and 8\farcs7$\times$8\farcs2, respectively. We used the \textsc{Aegean} software to create these point source catalogues, and together with the previously published \ac{MeerKAT} point source catalogue, we estimated spectral indices for the whole population of radio point sources in common. By cross-matching our \ac{ASKAP} catalogues with the \ac{MeerKAT} catalogue, we found 21,442 and 12,654 point sources in common for 944\,MHz and 1367\,MHz, respectively, within a 2\arcsec\ region. This point source catalogue will help to further our knowledge of the \ac{SMC} and highlights the power of the new generation of telescopes such as \ac{ASKAP} in studying different galactic populations.}


\keywords{Small Magellanic Cloud -- Catalogues -- ASKAP -- Radio-Continuum}

\end{strip}


\acresetall

\section{Introduction}
\label{sec:intro}

Over the past 50 years, advances in radio astronomy have improved our knowledge and understanding of the \ac{SMC}, a dwarf galaxy orbiting the Milky Way. The \ac{SMC} is located at a distance of $62.44 \pm 0.47$\,kpc, and has a considerable depth along our line of sight of up to 7\,kpc~\citep{2005MNRAS.357..304H}. Recent radio continuum surveys have been conducted at different frequencies, including radio surveys by telescopes such as \ac{MeerKAT}~\citep{2024MNRAS.529.2443C}, the \ac{ASKAP}~\citep{2019MNRAS.490.1202J} and the \ac{MWA}~\citep{2018MNRAS.480.2743F}. These allow for analysis into how the \ac{ISM} evolves, as well as shedding light on the properties and movements of sources such as \acp{SNR},  \acp{PN}, \acp{YSO}, and \HII\,regions~\citep{Filipovic2021}. 

The first published radio source catalogue of the \ac{SMC} was released in the 1970s by~\citet{1976MNRAS.174..393C,1996ASPC..112...91F}, which initiated a series of comprehensive radio astronomy studies. There have been several surveys conducted since then~\citep{ 1976AuJPh..29..329M, 1990PKS...C......0W, 1997A&AS..121..321F, 1998PASA...15..280T, 1998A&AS..130..421F, 2002MNRAS.335.1085F, 2004MNRAS.355...44P, 2005MNRAS.364..217F, 2006MNRAS.367.1379R, 2007MNRAS.376.1793P, 2011SerAJ.182...43W, 2011SerAJ.183...95C, 2011SerAJ.183..103W, 2012SerAJ.184...93W, 2012SerAJ.185...53W, 2018MNRAS.480.2743F,
2022PASA...39...34D,
2022PASA...39....5P,
2024ApJ...962..120M,
2024MNRAS.529.2443C}, which have gradually increased our knowledge of the dynamics and structure of the \ac{SMC}. 

These surveys have provided catalogues of \acp{SNR}, helping to identify their interactions with the surrounding interstellar medium and refine their distribution across the \ac{SMC}~\citep{2005MNRAS.364..217F, 2007MNRAS.376.1793P,2022MNRAS.513.1154M}. Furthermore, the identification and classification of \acp{PN} have improved our understanding of their evolution and chemical composition in the context of a dwarf galaxy~\citep{1996ASPC..112...91F,1997A&AS..121..321F}.

Recent radio-continuum surveys using new-generation interferometers such as \ac{ASKAP}, \ac{MeerKAT}, and \ac{MWA} have revolutionised our ability to study numerous Galactic and extragalactic radio sources and populations. These include discoveries and analyses of several objects, including Galactic and \ac{MC} \acp{SNR}~\citep{Kothes2017, Filipovic2023, BurgerSchiedlin2024, Khabibullin2024, 2024MNRAS.534.2918S, 2025ApJ...988...75B, 2025arXiv250504041F, 2025arXiv250615067S} and \ac{SNR} candidates~\citep{Lazarevic2024, Smeaton2024}, as well as the discovery of the first intergalactic \ac{SNR} J0624$-$6948~\citep{Filipovic2022, 2025A&A...693L..15S}, and the study of interesting Galactic and extragalactic sources such as \acp{RN}~\citep{2025PASA...42...32B}, Wolf-Rayet nebulae \citep{2025PASA...42..101B}, \acp{PWN}~\citep{2024PASA...41...32L, 2025MNRAS.537.2868A}, \ac{AGN}~\citep{2022MNRAS.516.1865V}, and \acp{ORC}~\citep{Norris2021ORC, 10.1093/mnrasl/slab041, Gupta2022, 2022MNRAS.513.1300N}. These surveys have also advanced our understanding of the evolution of \ac{AGN}, and star formation processes in low-metallicity environments~\citep{2021PASA...38....9H,2022MNRAS.510.3389U,2024MNRAS.529.2443C}. These large-area, high-sensitivity surveys allow for statistical characterisation of compact and extended sources across a range of physical scales, providing insights into the life cycles of galaxies, the role of magnetic fields, and the interaction between massive stars and the ISM~\citep{2022NatAs...6..828C}.

The radio-continuum catalogue of the \ac{SMC} presented by~\citet{2019MNRAS.490.1202J} was based on \ac{ASKAP} early science data taken at 1320\,MHz as part of the EMU pilot survey. While the survey was a significant step forward, the dataset was limited by early beam calibration models, reduced uv-coverage, and preliminary imaging and source extraction pipelines. As a result, the positional and flux accuracy, as well as the catalogue completeness, were subject to systematic limitations~\citep{2021PASA...38....9H}.

This paper presents a new \ac{ASKAP} radio-continuum survey and its corresponding catalogue. We carry out a comparative analysis with the recently published \ac{MeerKAT} survey of the \ac{SMC} by~\citet{2024MNRAS.529.2443C}, leveraging the extensive dataset to validate and assess the quality of our results.

In contrast, the new catalogue presented in this paper is based on \ac{ASKAP} observations at 944 and 1367\,MHz, using the latest imaging and calibration techniques from the ASKAPSoft pipeline~\citep{2019ascl.soft12003G}. The data benefit from improved primary beam models, better dynamic range, and deeper integrations. 
This catalogue represents an improvement over the earlier work of~\citet{2019MNRAS.490.1202J}, offering enhanced positional accuracy and more reliable flux measurements.

The structure of this paper is as follows: Section~\ref{data} describes \ac{ASKAP}'s observations, data, and processing. Section~\ref{Section 3} describes \ac{ASKAP}'s point source catalogue generating methodology. Section~\ref{Section 4} presents the results of the comparison between the \ac{ASKAP} and \ac{MeerKAT} catalogues. Section~\ref{Section 5} discusses these results, and Section~\ref{Section 6} provides the main conclusions.


\begin{figure*} 
    \centering
    \includegraphics[width=1\linewidth]{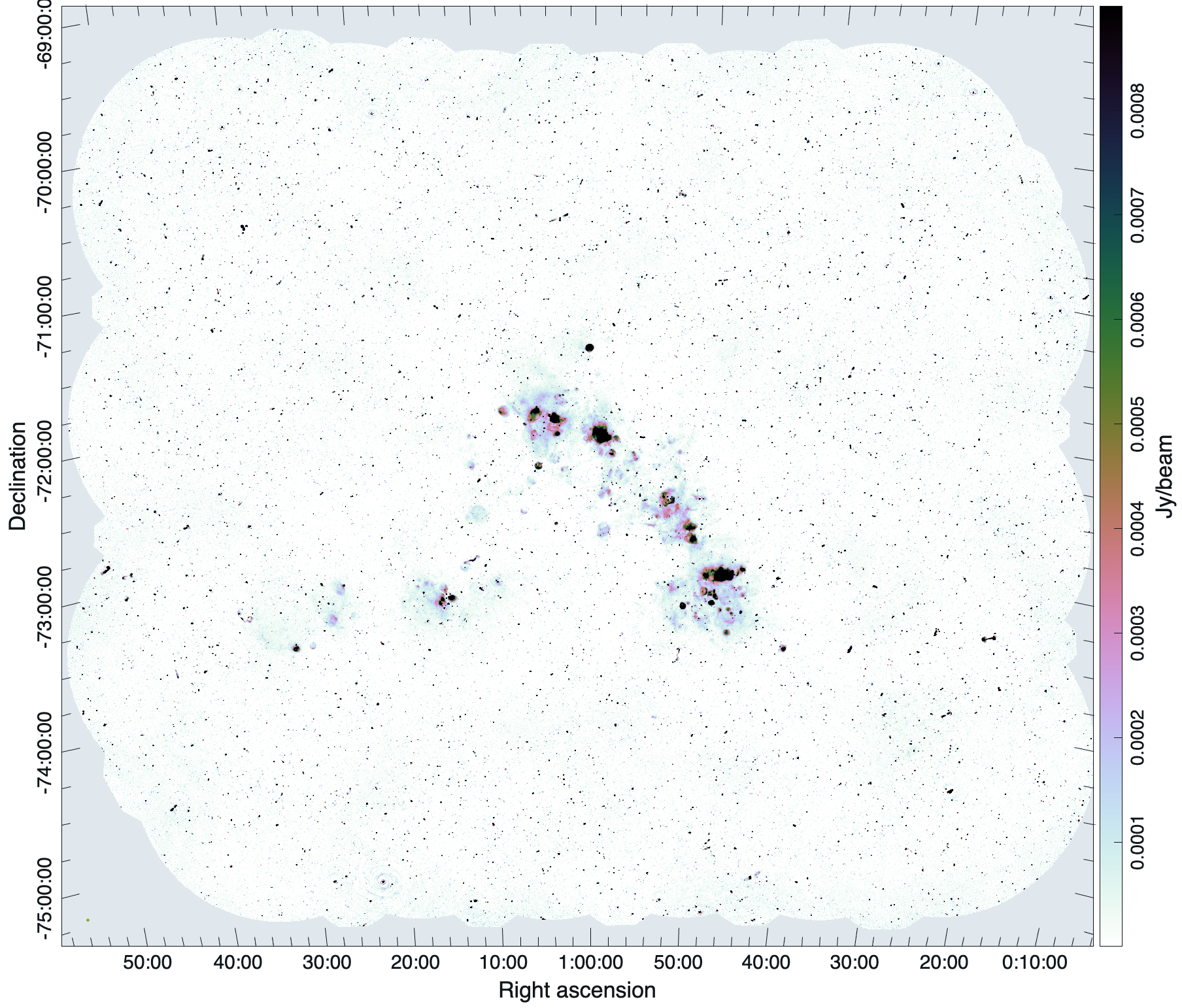}
    \caption{\ac{ASKAP} image of the \ac{SMC} at 944\,MHz. The beam size is $14\farcs5\times12\farcs2$ and the linearly scaled colour bar represents the image scale intensity range. The background noise level in the image is 26\,$\mu$Jy beam$^{-1}$. Coordinates are in the epoch J2000.}
    \label{fig:SMC_IMAGE_944 MHZ}
\end{figure*}

\begin{figure*}
    \centering
    \includegraphics[width=1\linewidth]{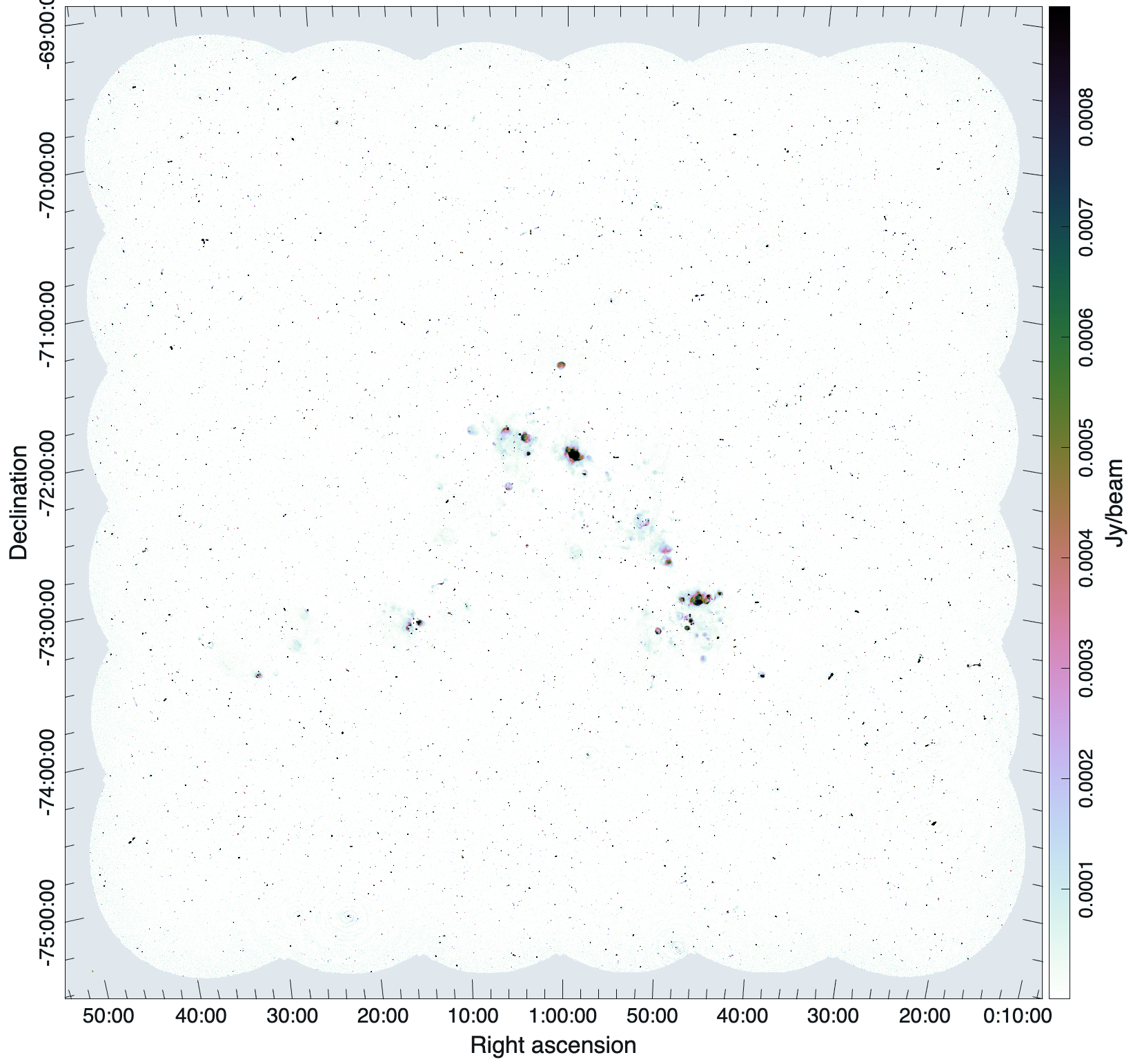}
    \caption{ASKAP  image of the \ac{SMC} at 1367\,MHz. The beam size is 8\farcs7$\times8\farcs2$ and the linearly scaled colour bar represents the image scale intensity range. The background noise level in the image is 28\,$\mu$Jy beam$^{-1}$. Coordinates are in the epoch J2000.}
    \label{fig:SMC image 1367}
\end{figure*}


\section{Data, Observation and Processing}
\label{data}

In this section, we present the data and images obtained with \ac{ASKAP} at 944 and 1367\,MHz~\citep{2016PASA...33...42M,2025PASA...42...71H}.

The \ac{ASKAP} Pilot Survey for POSSUM project code for both surveys is AS103\footnote{\url{http://hdl.handle.net/102.100.100/326207?index=1}} \citep{2025PASA...42...91G}. Observations were conducted on 2022-08-07 and 2022-09-18 using the \ac{ASKAP} array. The data corresponds with scheduling blocks SB43237 \url{(tile POSSUM\_0101-72)} and SB44127 \url{(tile POSSUM\_0101-72A)}. Observations were made at central frequencies of 944 and 1367\,MHz, with corresponding bandwidths of 288\,MHz. The beam sizes for the corresponding images are $14\farcs5\times12\farcs2$ and $8\farcs7\times8\farcs2$, respectively. The average background noise for the images are 26\,$\mu$Jy beam$^{-1}$ and 28 $\mu$Jy beam$^{-1}$, respectively (see Table~\ref{data_summary}). 

The data is accessible through the \ac{CSIRO} \ac{ASKAP} Science Data Archive (CASDA)\footnote{\url{https://data.csiro.au/domain/casda}}~\citep{2015IAUGA..2232458C, 2017ASPC..512...73C, 2020ASPC..522..263H}. 

The data reduction was performed with \textsc{ASKAPsoft} v1.7.1, using the ASKAP pipeline v1.9.5 for the 944 MHz and v1.9.7 for the 1367 MHz, which offers reliable calibration and imaging processes by utilising multi-frequency synthesis imaging and multi-scale cleaning methods~\citep{2019ascl.soft12003G}.

We show the full \ac{ASKAP} tiles of the \ac{SMC} field at the frequencies of 944\,MHz (Figure~\ref{fig:SMC_IMAGE_944 MHZ}) and 1367\,MHz (Figure~\ref{fig:SMC image 1367}). A summary of the \ac{ASKAP} observations is given in Table~\ref{data_summary}.

\begin{table*}
\centering
\caption{Summary of \ac{ASKAP} data and observations}
\label{data_summary}
\begin{tabular}{p{6cm} c c}
\hline
\textbf{Frequency (MHZ)} & \textbf{944\,MHz} & \textbf{1367\,MHz} \\ \hline
\textbf{Project Code} & \multicolumn{2}{c}{AS103} \\ 
\textbf{Bandwidth (MHz)} & \multicolumn{2}{c}{288} \\ 
\textbf{Stokes Parameter} & \multicolumn{2}{c}{STOKES I} \\ 
\textbf{Observation Date} & 2022-08-07 & 2022-09-18 \\ 
\textbf{Scheduling Block (SBID)} & SB43237  & SB44127  \\ 
\textbf{Beam Size (\arcsec$\times$\arcsec)} & 14.5 $\times$ 12.2 & 8.7 $\times$ 8.2 \\ 
\textbf{RMS ($\mu$Jy beam$^{-1}$)} & 26 & 30 \\ \hline
\end{tabular}
\end{table*}


\begin{figure*}
    \centering
    \includegraphics[width=\textwidth]{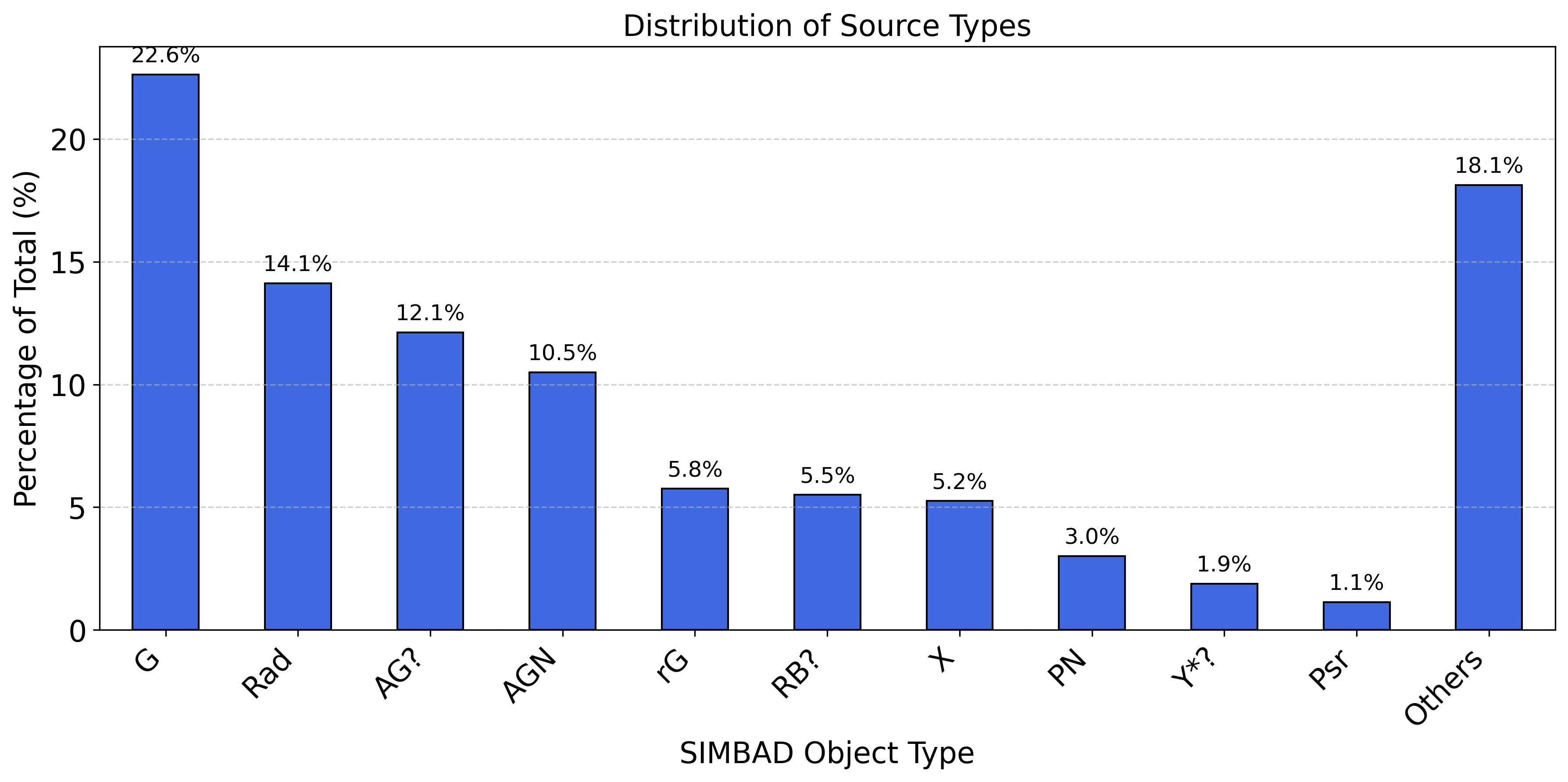}
    \caption{Distribution of SIMBAD object types from cross-matched sources.
G = galaxy; Rad = radio source; AG? = possible active galaxy; AGN = active galactic nucleus; rG = radio galaxy; RB? = possible radio burst source; X = X-ray source; PN = planetary nebula; Y*? = possible young stellar object; Psr = pulsar; Others = rare types.}
    \label{fig:simbad_histogram}
\end{figure*}

\begin{table*} 
\small 
\begin{center} 
\caption{Example of the \ac{ASKAP} 944\,MHz and 1367\,MHz point source catalogue of 45,886 objects in the directions of the \ac{SMC} with their positions, peak and integrated flux densities. The columns provided are as follows: (1) Source name; (2 and 3) source position; (4 and 5) source peak and integrated flux density at 944\,MHz with associated uncertainty; (6 and 7) source peak and integrated flux density at 1367\,MHz with associated uncertainty; (8) spectral index ($\alpha$) fitted using all available frequency measurements (ASKAP 944/1367\,MHz and up to 12 MeerKAT sub-bands) with the associated error; (9) number of flux-density measurements used in the spectral index fit. The full catalogue is provided through the \textsc{VizieR} service and as supplementary material.}
\label{tab:ASKAP_944MHz_1367MHz_catalogue}  
\resizebox{\textwidth}{!}{
\begin{tabular}{lllcccccccc}  
\hline  
\textbf{Name} & \textbf{RA (J2000)} & \textbf{Dec (J2000)} & \textbf{S$_{\rm peak\,944\,MHz}$} & \textbf{S$_{\rm int\,944\,MHz}$} & \textbf{S$_{\rm peak\,1367\,MHz}$} & \textbf{S$_{\rm int\,1367\,MHz}$} & \textbf{$\alpha\pm\Delta\alpha$} & \textbf{n} \\  
& & & \textbf{(mJy~beam$^{-1}$)} & \textbf{(mJy)} & \textbf{(mJy~beam$^{-1}$)} & \textbf{(mJy)} & & \\  
\hline  
J003334--703819 & 00:33:34 & --70:38:19 & 0.168$\pm$0.008 & 0.168$\pm$0.008 & 0.171$\pm$0.009 & 0.176$\pm$0.009 & --0.36$\pm$0.46 & 7 \\  
J002906--745338 & 00:29:06 & --74:53:38 & 0.537$\pm$0.027 & 0.537$\pm$0.027 & 0.420$\pm$0.021 & 0.420$\pm$0.021 & --0.98$\pm$0.089 & 14 \\  
J004938--741929 & 00:49:38 & --74:19:29 & 0.260$\pm$0.013 & 0.354$\pm$0.018 & 0.135$\pm$0.007 & 0.135$\pm$0.007 & --1.68$\pm$0.31 & 12 \\  
J005535--722836 & 00:55:35 & --72:28:36 & 0.733$\pm$0.037 & 0.747$\pm$0.037 & 0.437$\pm$0.022 & 0.565$\pm$0.028 & --0.90$\pm$0.11 & 14 \\  
J005701--715359 & 00:57:01 & --71:53:59 & 0.494$\pm$0.025 & 0.509$\pm$0.025 & 0.297$\pm$0.015 & 0.390$\pm$0.020 & --0.87$\pm$0.14 & 14 \\  
J005833--723212 & 00:58:33 & --72:32:12 & 0.617$\pm$0.031 & 0.617$\pm$0.031 & 0.683$\pm$0.034 & 0.733$\pm$0.037 & 0.17$\pm$0.08 & 14 \\  
J005617--704541 & 00:56:17 & --70:45:41 & 0.278$\pm$0.014 & 0.317$\pm$0.016 & 0.297$\pm$0.015 & 0.297$\pm$0.015 & --0.45$\pm$0.18 & 12 \\  
J011031--715711 & 01:10:31 & --71:57:11 & 0.234$\pm$0.012 & 0.234$\pm$0.012 & 0.146$\pm$0.007 & 0.159$\pm$0.008 & --0.84$\pm$0.21 & 12 \\  
J013559--713914 & 01:35:59 & --71:39:14 & 0.339$\pm$0.017 & 0.339$\pm$0.017 & 0.206$\pm$0.010 & 0.212$\pm$0.011 & --0.86$\pm$0.26 & 13 \\  
J004538--734546 & 00:45:38 & --73:45:46 & 0.189$\pm$0.009 & 0.189$\pm$0.009 & 0.205$\pm$0.010 & 0.205$\pm$0.010 & --0.30$\pm$0.22 & 11 \\  
J002716--695752 & 00:27:16 & --69:57:52 & 0.443$\pm$0.022 & 0.452$\pm$0.023 & 0.393$\pm$0.020 & 0.426$\pm$0.021 & --0.81$\pm$0.24 & 14 \\  
J013215--740949 & 01:32:15 & --74:09:49 & 0.200$\pm$0.010 & 0.200$\pm$0.010 & 0.134$\pm$0.007 & 0.134$\pm$0.007 & --0.87$\pm$0.36 & 12 \\  
J013629--703303 & 01:36:29 & --70:33:03 & 1.093$\pm$0.055 & 1.138$\pm$0.057 & 0.644$\pm$0.032 & 0.756$\pm$0.038 & --0.48$\pm$0.11 & 14 \\  
J013121--711130 & 01:31:21 & --71:11:30 & 2.362$\pm$0.118 & 2.472$\pm$0.124 & 1.961$\pm$0.098 & 2.083$\pm$0.104 & --0.41$\pm$0.049 & 14 \\  
J011312--711956 & 01:13:12 & --71:19:56 & 0.153$\pm$0.008 & 0.172$\pm$0.009 & 0.161$\pm$0.008 & 0.189$\pm$0.009 & --0.57$\pm$0.33 & 12 \\  
\hline  
\end{tabular}  
}
\end{center}  
\end{table*}  

\section{Point Source Catalogue Generation Methodology}
\label{Section 3}
This section presents the systematic method undertaken to generate the point source catalogue that we adopt in our analysis. This process was optimised to achieve a maximum degree of reliability, completeness, and consistency across the entire observed field. We present the step-by-step data processing and source detection steps, from first source detection to post-processing filters and verification, employed to calibrate the source catalogue and to reduce noise caused by artifacts, extended sources, and noise-limited regions. This process employs automated source-finding programs together with statistically warranted quality assessment criteria and final visual checking, in compliance with best practices in current broad field radio-continuum surveys.

\vspace{0.2cm}
\textbf{Step 1: Initial source detection and catalogue Generation}
\vspace{0.1cm}

The source finding program \textsc{Aegean}\footnote{\url{https://github.com/PaulHancock/Aegean}}~\citep{2012MNRAS.422.1812H,2018PASA...35...11H} was employed to generate initial source lists from the \ac{ASKAP} images at a 3-sigma detection threshold. These images, which exhibit varying noise levels due to the presence of multiple beams and bright source artifacts, were first processed using the \ac{BANE}\footnote{\url{https://github.com/PaulHancock/Aegean/blob/main/AegeanTools.py}} routine. BANE generates \ac{RMS} background noise maps using a sliding box-car and sigma-clipping algorithm~\citep{2018PASA...35...11H}. The initial catalogue included all sources detected at the 3-sigma level with default parameters.

\vspace{0.2cm}
\textbf{Step 2: Masking Sources Based on FLAGS and SOURCES}
\vspace{0.1cm}

We excluded sources with \texttt{FLAGS} not equal to 0 or \texttt{SOURCES} less than 1, following recommendations in the \textsc{Aegean} documentation\footnote{\url{https://github.com/PaulHancock/Aegean/blob/main/doc/includes/aegean.md}}. According to the documentation, non-zero \texttt{FLAGS} indicate unreliable fits, while a source count (\texttt{SOURCES}) below 1 suggests inadequate detection quality. This step was crucial for ensuring that only well-fit, reliable sources were retained in the final catalogue.

\vspace{0.2cm}
\textbf{Step 3: Axis Ratio Masking}
\vspace{0.1cm}

To exclude extended or poorly resolved sources, we masked sources based on their major and minor axis ratios. The major axis ratio (\( R_a \)) and minor axis ratio (\( R_b \)) were calculated as follows:
\[
R_a = \frac{a}{\text{psf}_a}, \quad R_b = \frac{b}{\text{psf}_b},
\]
where \(a\) and \(b\) are the source's measured major and minor axes, and \(\text{psf}_a\) and \(\text{psf}_b\) are the corresponding beam's major and minor axes. A perfect point source should have an angular size equal to the beam size, meaning \( R_a = 1 \) and \( R_b = 1 \). Any source with an angular size larger than the beam is considered resolved, while those smaller than the beam are unresolved and handled separately using the peak flux density (\( S_{\rm peak} \)) instead of the integrated flux density (\( S_{\rm int} \)).

To retain compact sources, we excluded all sources with \(R_a > 1.2\) or \(R_b > 1.2\), corresponding to 20\% larger than the beam size. This criterion ensures that we remove significantly extended sources while preserving compact and point-like sources.

At the SMC distance of $62.44\pm0.47$\,kpc, $1\arcsec\simeq0.303$\,pc. Thus, the ASKAP beams correspond to physical scales of $\sim14.5\arcsec\times12.2\arcsec\approx4.39\times3.69$\,pc at 944\,MHz and $\sim8.7\arcsec\times8.2\arcsec\approx2.63\times2.48$\,pc at 1367\,MHz. Our $1.2\times$ beam-size threshold therefore translates to excluding sources larger than $\sim17.4\arcsec\times14.6\arcsec\approx5.27\times4.43$\,pc (944\,MHz) and $\sim10.44\arcsec\times9.84\arcsec\approx3.16\times2.98$\,pc (1367\,MHz).

This selection method may introduce minor biases, such as the inclusion of faint extended sources near the threshold and the exclusion of marginally resolved sources due to beam-fitting variations. Despite these factors, the 1.2 threshold effectively balances the rejection of extended sources while preserving a reliable sample of compact radio sources.

\vspace{0.2cm}
\textbf{Step 4: Correcting Integrated Flux Values}
\vspace{0.1cm}

Some sources exhibited integrated flux densities lower than their peak flux densities. When a source extractor provides such results, it means that the source is unresolved, and the integrated flux density should be equal to the peak flux density. To correct these values, we applied the following logical operation:
\[
S_{\text{int}} = 
\begin{cases} 
S_{\text{peak}}, & \text{if } S_{\text{int}} < S_{\text{peak}}, \\
S_{\text{int}}, & \text{otherwise.}
\end{cases}
\]
This correction ensured that all unresolved sources in the catalogue were properly characterised. The corrected catalogue is shown in Table~\ref{tab:ASKAP_944MHz_1367MHz_catalogue}, where integrated flux densities have been adjusted.

\vspace{0.2cm}
\textbf{Step 5: Local RMS Ratio Masking}
\vspace{0.1cm}

To minimize contamination from high-noise regions, we removed sources located in areas where the local RMS values exceeded 2.5-sigma (65~\textmu Jy) of the background RMS noise. This threshold was chosen to balance the exclusion of sources affected by elevated background noise while retaining genuine detections. By applying this criterion, we ensured the removal of point sources with unreliable flux measurements due to high local noise levels.

\vspace{0.2cm}
\textbf{Step 6: Visual Inspection and Validation}
\vspace{0.1cm}

Following the masking steps, the remaining sources were visually inspected to ensure the validity of the detections, similar to the methodologies employed by~\citet{2019MNRAS.490.1202J} and~\citet{2024MNRAS.529.2443C}. This step helped identify and remove any remaining artifacts, including those caused by \ac{BCE} sources that had been split into multiple detections.

\vspace{0.2cm}
\textbf{Step 7: Merging catalogues with different frequencies}
\vspace{0.1cm}

To validate the quality and reliability of our \ac{ASKAP} 944\,MHz and 1367\,MHz catalogues of the \ac{SMC}, we performed a comprehensive comparison with the MeerKAT 1283\,MHz catalogue \citep{2024MNRAS.529.2443C} and the Milliquas catalogue \citep{2023OJAp....6E..49F}. Using TOPCAT\footnote{\url{http://www.starlink.ac.uk/topcat/}}~\citep{2005ASPC..347...29T}, we crossmatched the sources in our \ac{ASKAP} catalogues with those in the MeerKAT catalogue within a 2\arcsec\ region, ensuring accurate source alignment given the known calibration issues~\citep{2024MNRAS.529.2443C}. In this merging process, we examined the positional offsets between our \ac{ASKAP} catalogues and the MeerKAT catalogue to identify any offsets or variations (see Section \ref{Results-PO}). Additionally, we assessed the flux densities of the matched sources to verify the consistency and reliability of the measurements across different instruments and frequencies (see Section \ref{Results-FC}). These multi-frequency flux comparisons also enabled the estimation of spectral indices, providing insight into the emission properties of radio sources across the \ac{SMC} field (see Section \ref{Results-SI}). To assess the reliability of the spectral index fits, we computed reduced chi-squared ($\chi^2_\nu$) values for each source (see Figure~\ref{fig:Figure 9}).

A significant fraction of fits showed elevated reduced $\chi^2_\nu$ values, which can have numerous causes. We found that these elevated levels corresponded with sources that had significantly underestimated flux density uncertainties, a known issue with the Aegean program \citep{2018PASA...35...11H}. We therefore adopted a conservative error model for sources with reduced $\chi^2_\nu \gg 1$, assuming a 5\% flux density uncertainty for sources with $S_\nu > 1$\,mJy and 10\% for those with $S_\nu < 1$\,mJy (see Figure~\ref{fig:Figure 10}). We find that this method improves the spectral index fits and removes this fraction of elevated $\chi^2_\nu$ values, indicating that the issue arose from underestimated uncertainties rather than poor model behaviour. For sources with $\chi^2_\nu < 1$, the original flux density uncertainties from the catalogue were retained (see Figure~\ref{fig:Figure 9}). This selective adjustment provided a more realistic error estimation and improved the statistical reliability of the spectral index fits. These analyses build on previous \ac{ASKAP} surveys \citep{2019MNRAS.490.1202J} and provide a robust foundation for studying radio source populations in the \ac{SMC}, including galaxies, active galactic nuclei (AGN), and supernova remnants (SNRs). Results are presented in Section~\ref{Section 4}.

\vspace{0.2cm}
\textbf{Step 8: Cross-matching with SIMBAD}
\vspace{0.1cm}

As a final step, our catalogue was cross-matched with the \textsc{SIMBAD} database using the CDS X-Match service \citep{2012ASPC..461..291B} within a 2\arcsec\ region to identify known source types. This procedure yielded 800 matched sources. The distribution of these sources by object type is shown in Figure~\ref{fig:simbad_histogram}. The dominant populations are galaxies (G; 22.6\%), radio sources (Rad; 14.1\%), \ac{AGN} candidates (AG?; 12.1\%), and confirmed \acp{AGN} (10.5\%). Additional notable classes include radio galaxies (rG), radio binaries (RB?), and X-ray sources (X), each contributing between 5--6\% of the total. The remaining categories are less frequent types such as \acp{PN}, candidate \acp{YSO} (Y*?), and pulsars (Psr). All remaining classifications with individual contributions below 1\% are grouped under ``Others'' in the figure. This category includes a wide range of less common radio sources, including \acp{QSO}, emission-line galaxies (Em*), Seyfert galaxies (Sy1, Sy2, SyG), \textsc{H\,ii} regions (\textsc{H\,ii}), stellar clusters (Cl*), \acp{SN}, white dwarf candidates (WD?), and several classes of variable and evolved stars (e.g., LP*, RR*, dS*, and Mi*). This cross-matching step gives a valuable overview of the contents of the catalogue in terms of the different populations of the \ac{SMC}, including \acp{PN}, \acp{SNR}, radio stars, and background sources. Approximately 59\% of the 800 matched sources correspond to extragalactic sources located within the field of the \ac{SMC} region but lying beyond the \ac{SMC} itself.

\FloatBarrier

    


\begin{figure*}
    
\begin{center}
    \includegraphics[width=1\textwidth, keepaspectratio]{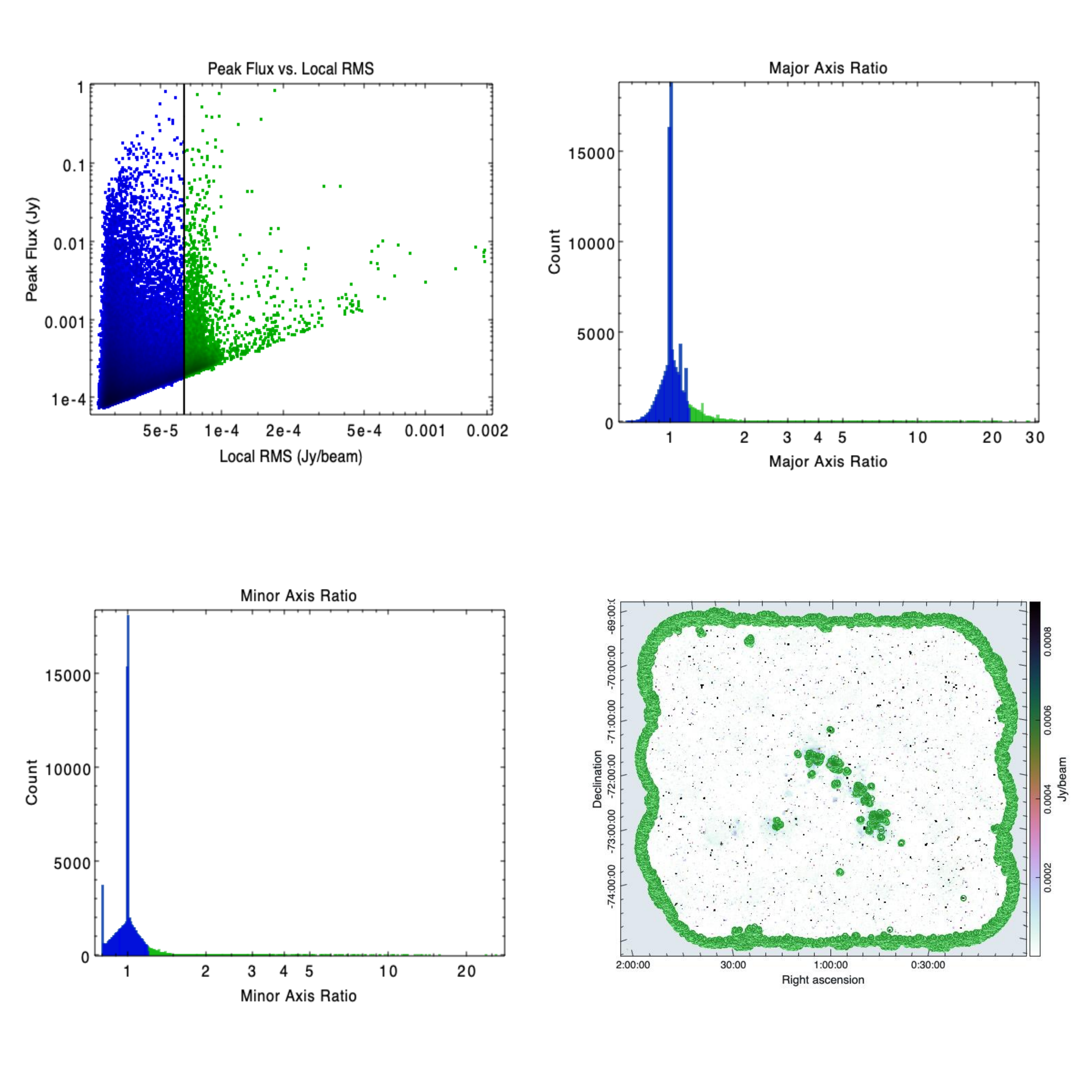}
\end{center}
  
    \caption{Overview of \ac{ASKAP} 944\,MHz analysis results. 
    \textbf{Top Left}: Peak flux as a function of RMS for \ac{ASKAP} 944\,MHz on a logarithmic scale. The grey line indicates the threshold at 65 $\mu$Jy beam$^{-1}$. Retained sources are in blue and discarded sources in green. 
    \textbf{Top Right}: Major Axis Ratio distribution for \ac{ASKAP} 944\,MHz sources. Retained sources are in blue, and discarded sources are in green. 
    \textbf{Bottom Left}: Minor Axis Ratio distribution for \ac{ASKAP} 944\,MHz sources. Retained sources are in blue, and discarded sources are in green. 
    \textbf{Bottom Right}: \ac{ASKAP} 944\,MHz image with green circles representing the masked sources, predominantly located at the edges of the image and in regions with \ac{BCE} sources.}

    \label{fig:Figure 3}
\end{figure*}

\begin{figure*}[p]
    \centering

    \begin{subfigure}[t]{0.49\textwidth}
        \centering
        \includegraphics[width=\linewidth]{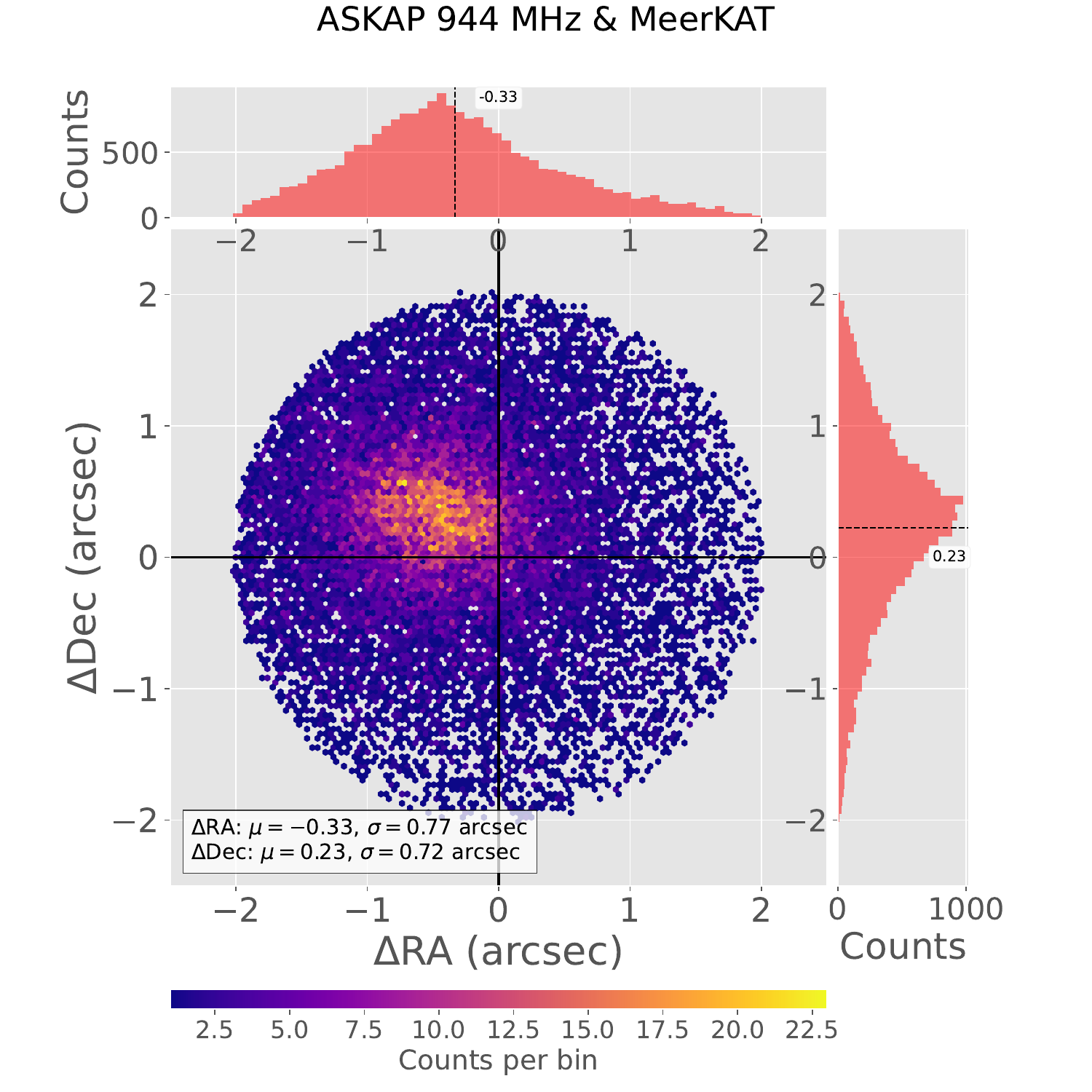}
        \label{fig:ASKAP944MKT}
    \end{subfigure}
    \hfill
    \begin{subfigure}[t]{0.49\textwidth}
        \centering
        \includegraphics[width=\linewidth]{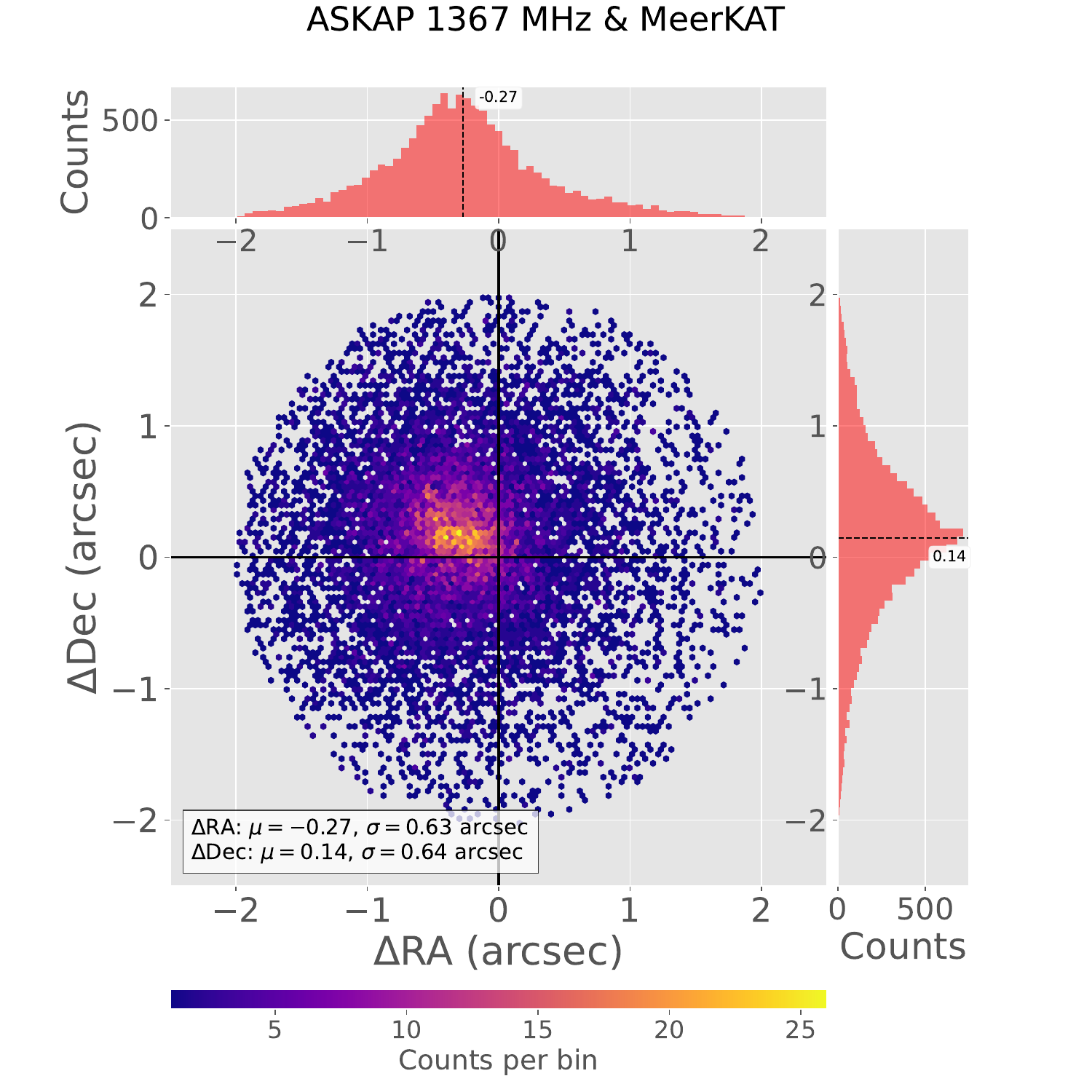}
        \label{fig:ASKAP1367MKT}
    \end{subfigure}

    \vspace{0.5cm}

    \begin{subfigure}[t]{0.49\textwidth}
        \centering
        \includegraphics[width=\linewidth]{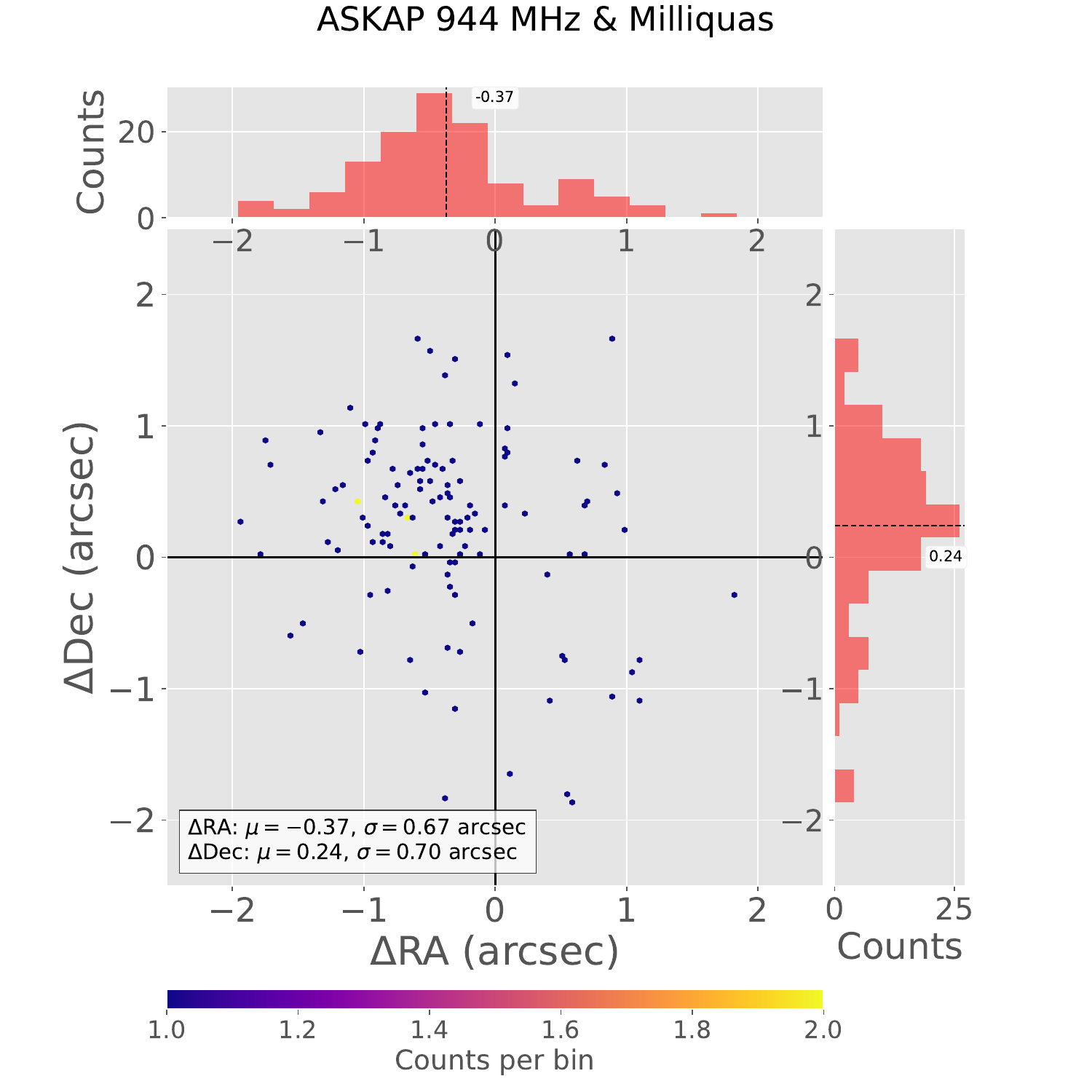}
        \label{fig:ASKAP944Milliquas}
    \end{subfigure}
    \hfill
    \begin{subfigure}[t]{0.49\textwidth}
        \centering
        \includegraphics[width=\linewidth]{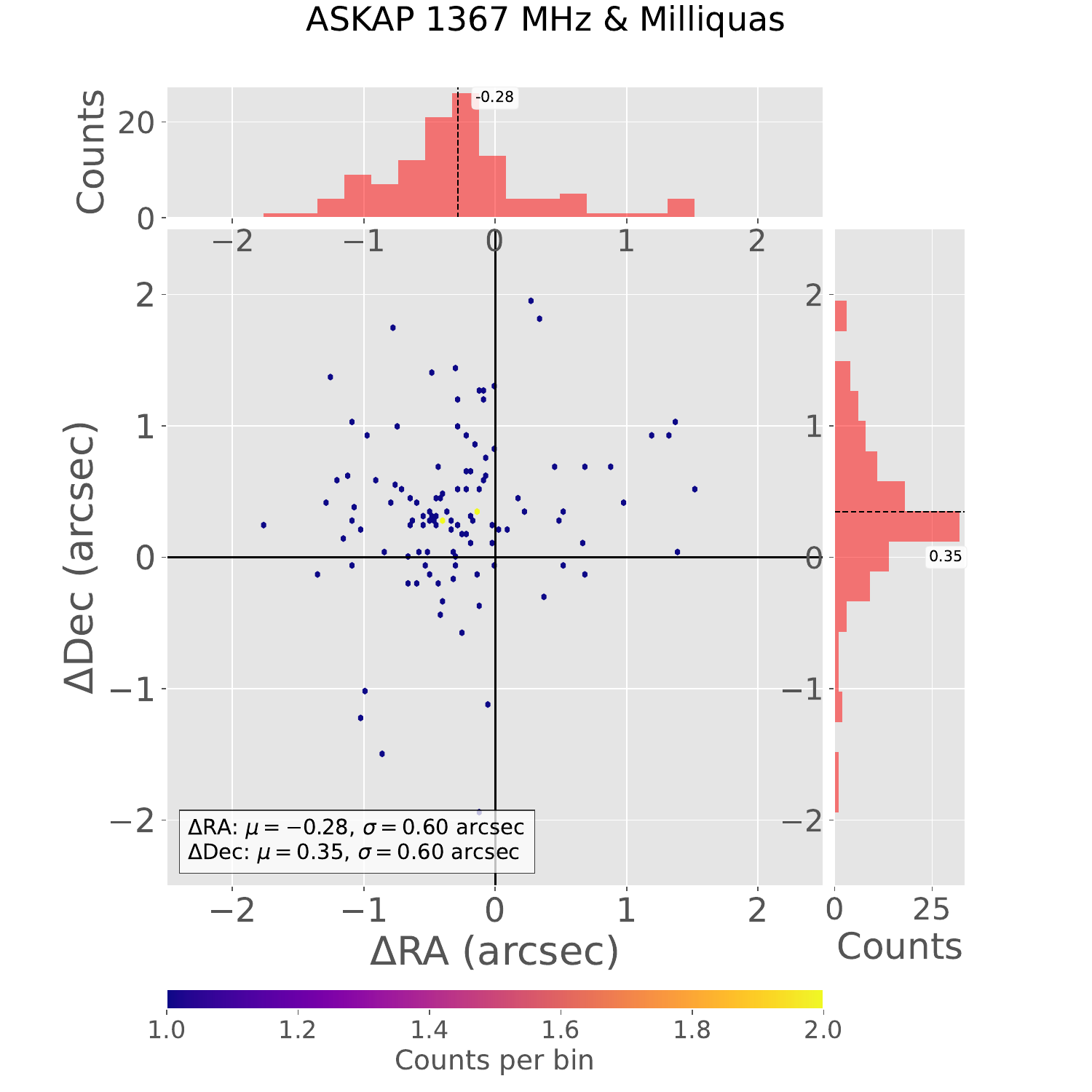}
        \label{fig:ASKAP1367Milliquas}
    \end{subfigure}

    \caption{Positional differences between \ac{ASKAP} and external catalogues, based on cross-matching at 2\arcsec\ radius. Each panel displays local density-colour-coded hexbin maps and marginal histograms. The black dashed lines in the red histograms represent the median values of the respective distributions. The points are colour-coded to indicate local density, from yellow for high density to purple for low density.}
    \label{fig:posdiff}
\end{figure*}

\begin{figure*}[p]
    \centering

    \begin{subfigure}[t]{0.45\textwidth}
        \centering
        \includegraphics[width=\linewidth]{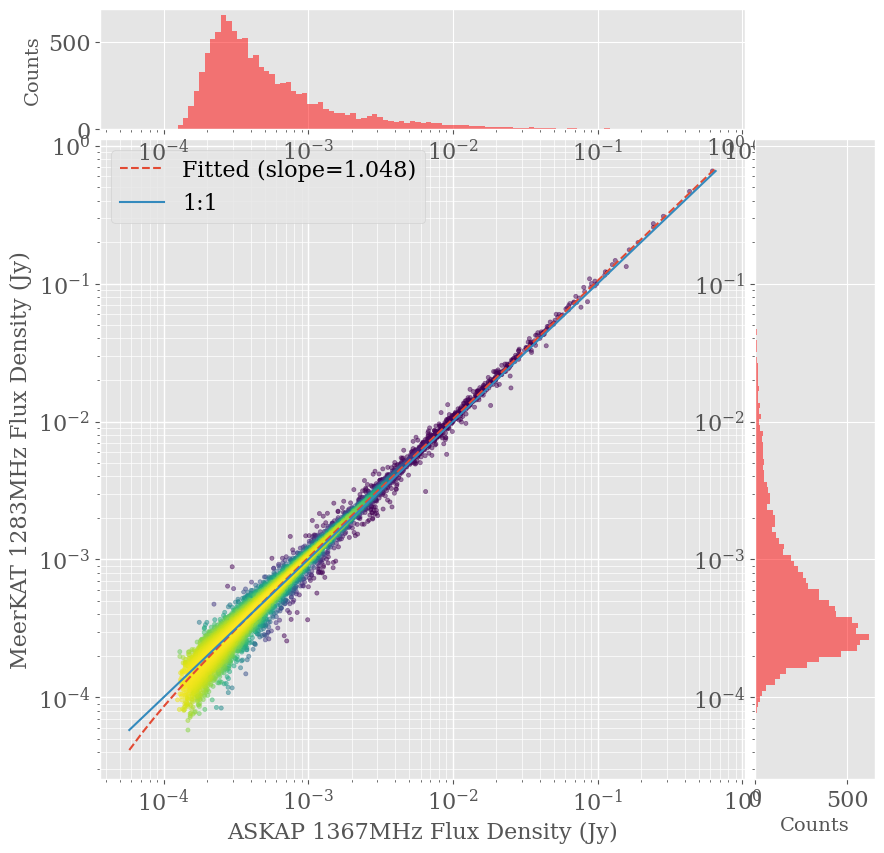}
        \label{fig:flux-comparison-5sigma}
    \end{subfigure}
    \hfill
    \begin{subfigure}[t]{0.45\textwidth}
        \centering
        \includegraphics[width=\linewidth]{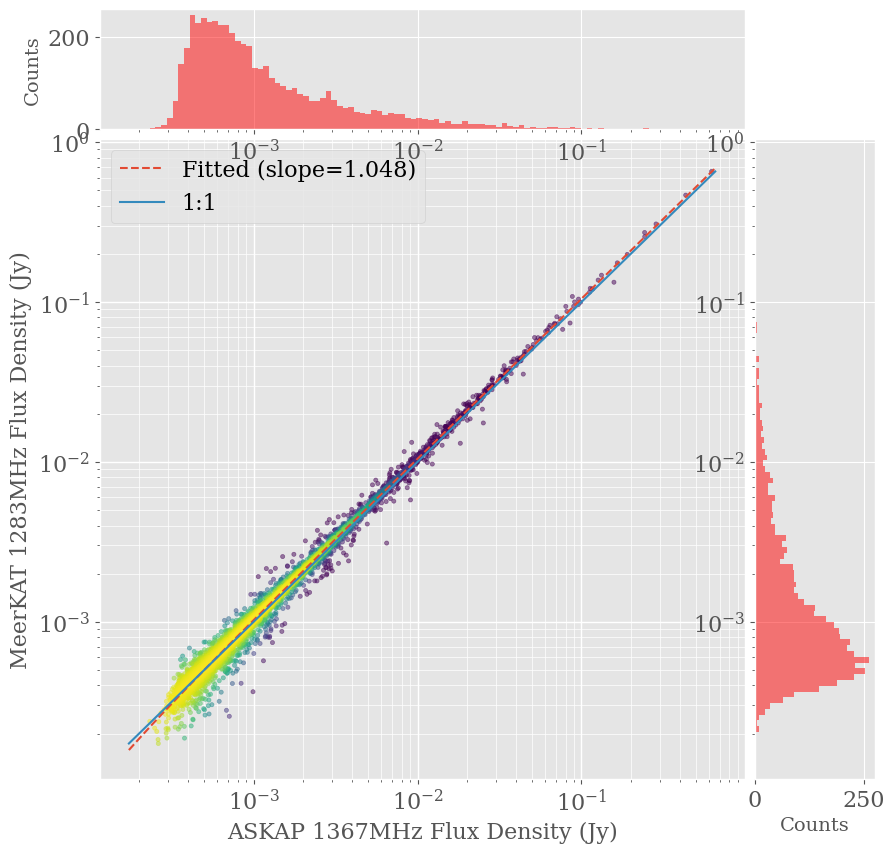}
        \label{fig:flux-comparison-10sigma}
    \end{subfigure}

    \vspace{0.7cm}

    \begin{subfigure}[t]{0.45\textwidth}
        \centering
        \includegraphics[width=\linewidth]{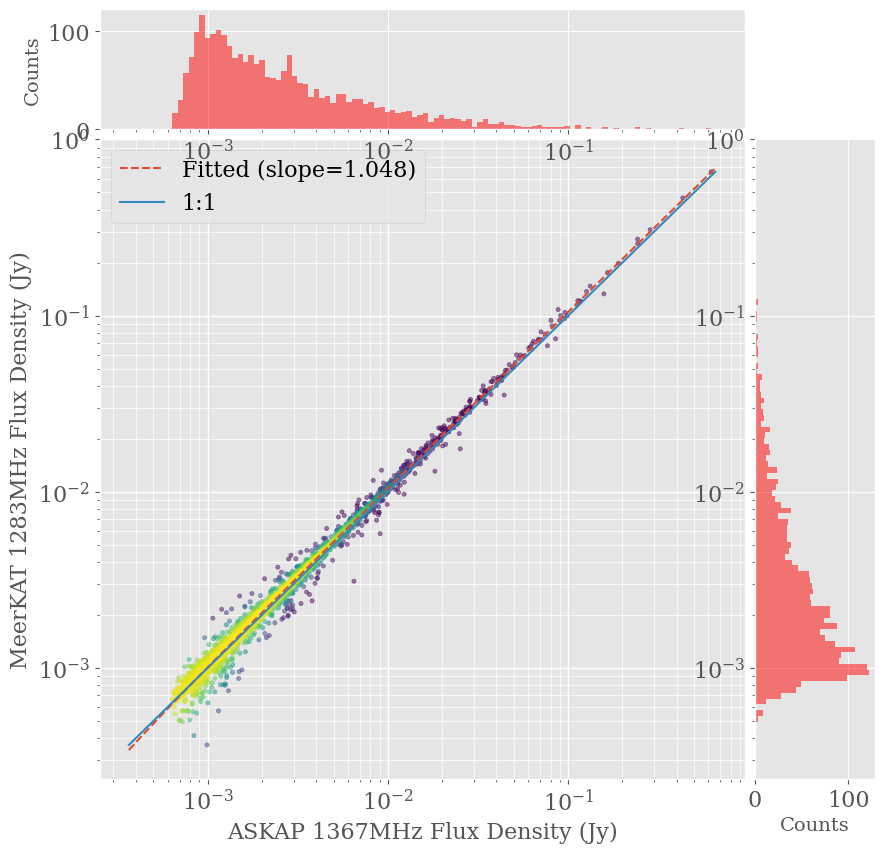}
        \label{fig:flux-comparison-20sigma}
    \end{subfigure}
    \hfill
    \begin{subfigure}[t]{0.45\textwidth}
        \centering
        \includegraphics[width=\linewidth]{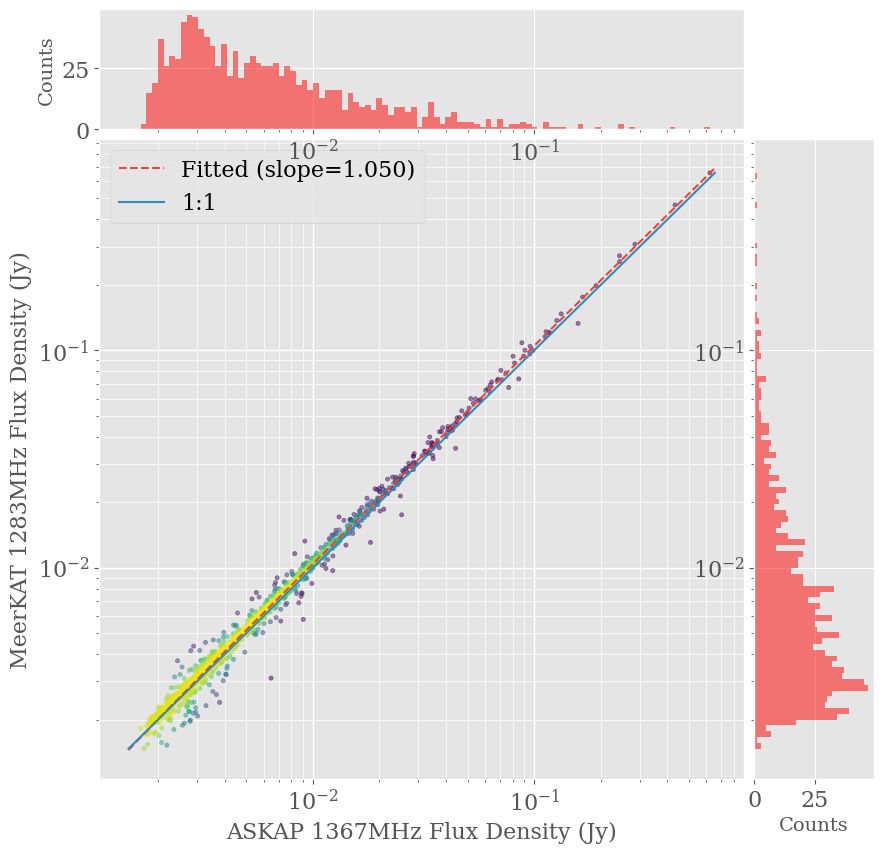}
        \label{fig:flux-comparison-50sigma}
    \end{subfigure}

    \caption{Flux density comparisons between \ac{ASKAP} 1367\,MHz and MeerKAT 1283\,MHz under varying signal-to-noise thresholds and visual encodings. The four panels correspond to catalogue matches for \(\geq 5\sigma\), \(\geq 10\sigma\), \(\geq 20\sigma\), and \(\geq 50\sigma\) sources, respectively. Sources are colour-coded by local density, with yellow indicating high density regions and purple indicating low density regions.}
    \label{fig:Figure 5}
\end{figure*}



\begin{figure*}
    
\begin{center}
    \includegraphics[width=1\textwidth, keepaspectratio]{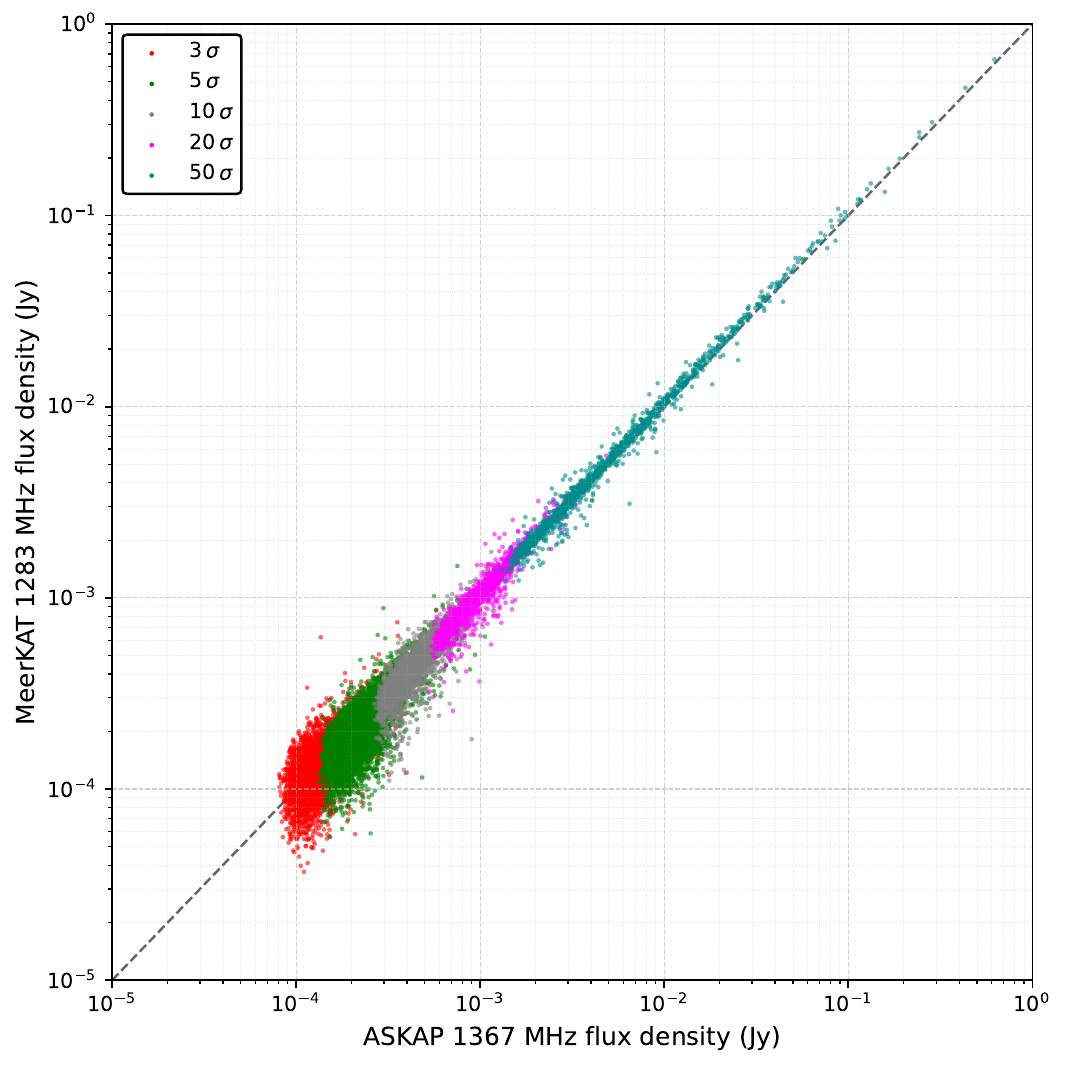}
\end{center}

\caption{ Flux density comparisons between \ac{ASKAP} 1367\,MHz and MeerKAT 1283\,MHz across all sources from $3\sigma$ to $50\sigma$, colour-coded by signal-to-noise ratio (S/N), ranging from $3\sigma$ (red) to $50\sigma$ (cyan). The fitted slope remains consistent across S/N bins, indicating robust flux agreement.} 

   \label{fig:Figure 6}

\end{figure*}




\begin{figure*}
    \centering
    \vspace{0.45cm}
    \includegraphics[width=0.95\textwidth]{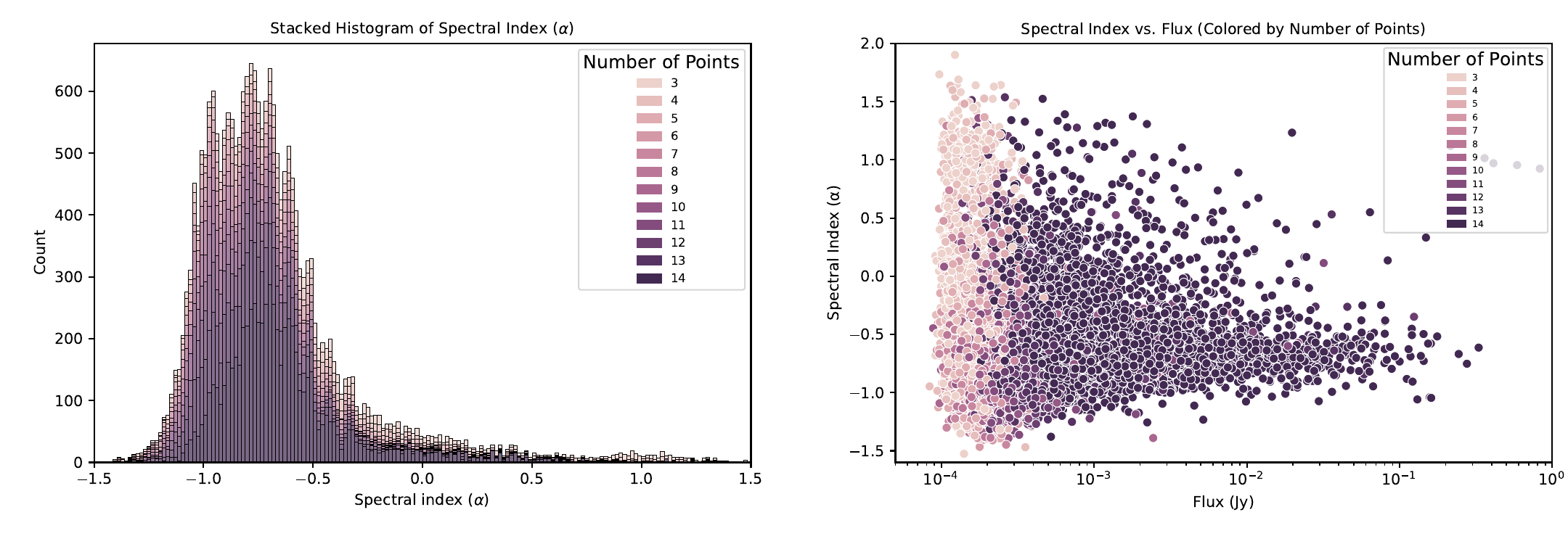}
    \vspace{0.45cm}
    \caption{Distribution of spectral indices across 14 different frequencies from \ac{ASKAP} and MeerKAT. {\bf Left}: Stacked histogram showing the point source spectral index distribution. The colours represent the number of channels used in the spectral index
fit. {\bf Right}: Scatter plot showing the distribution of spectral indices per integrated flux density. The colours represent the number of channels used in the spectral index fit. In both
figures, light colours represent small numbers, and dark colours represent high.}
    \label{fig:Figure 7}
\end{figure*}
    
    \vspace{0.45cm}



\begin{figure*}
    \centering
    \vspace{0.45cm}
    \includegraphics[width=0.95\textwidth]{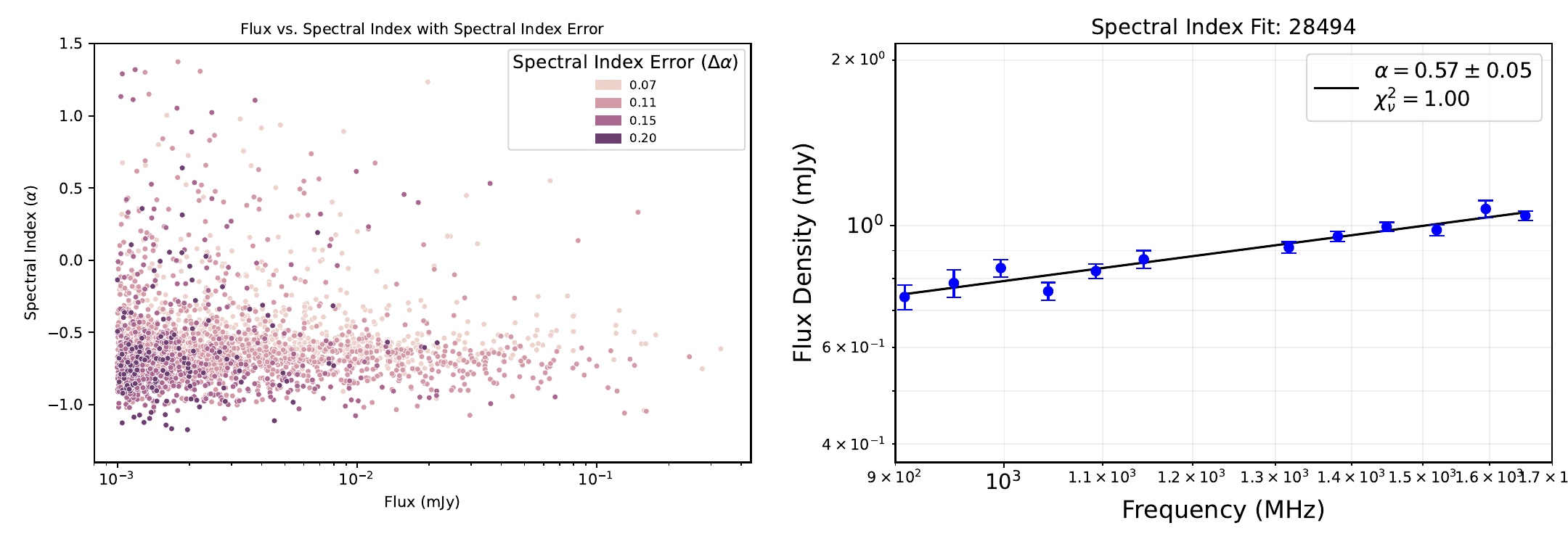}
    \vspace{0.45cm}
    
    \caption{{\bf Left}: The distribution of spectral indices per integrated flux densities for 2,726 sources. The sources were restricted to those with 14 frequency-flux measurements, flux density $>$1\,mJy, spectral index error $\Delta \alpha < 0.2$, and $\chi_\nu^2 < 1$. The colour bar represents the spectral index fit error. {\bf Right}: Example of a spectral index distribution of the source with \texttt{mainID 28494} ($\chi^2_{\nu} = 1$).Flux density is plotted as a function of frequency on a log--log scale. The solid black line represents the orthogonal distance regression (ODR) fit in log space, from which the spectral index ($\alpha$) and reduced chi-squared ($\chi^2_{\nu}$) were derived. Vertical error bars indicate the assumed flux density uncertainties. All 14 available frequency measurements were used in both fits. }
    
    \label{fig:Figure 8}
\end{figure*}

\begin{figure*}
    \centering
    \begin{minipage}{0.49\textwidth}
        \centering
        \includegraphics[trim={0 0 0 0},clip,width=\textwidth]{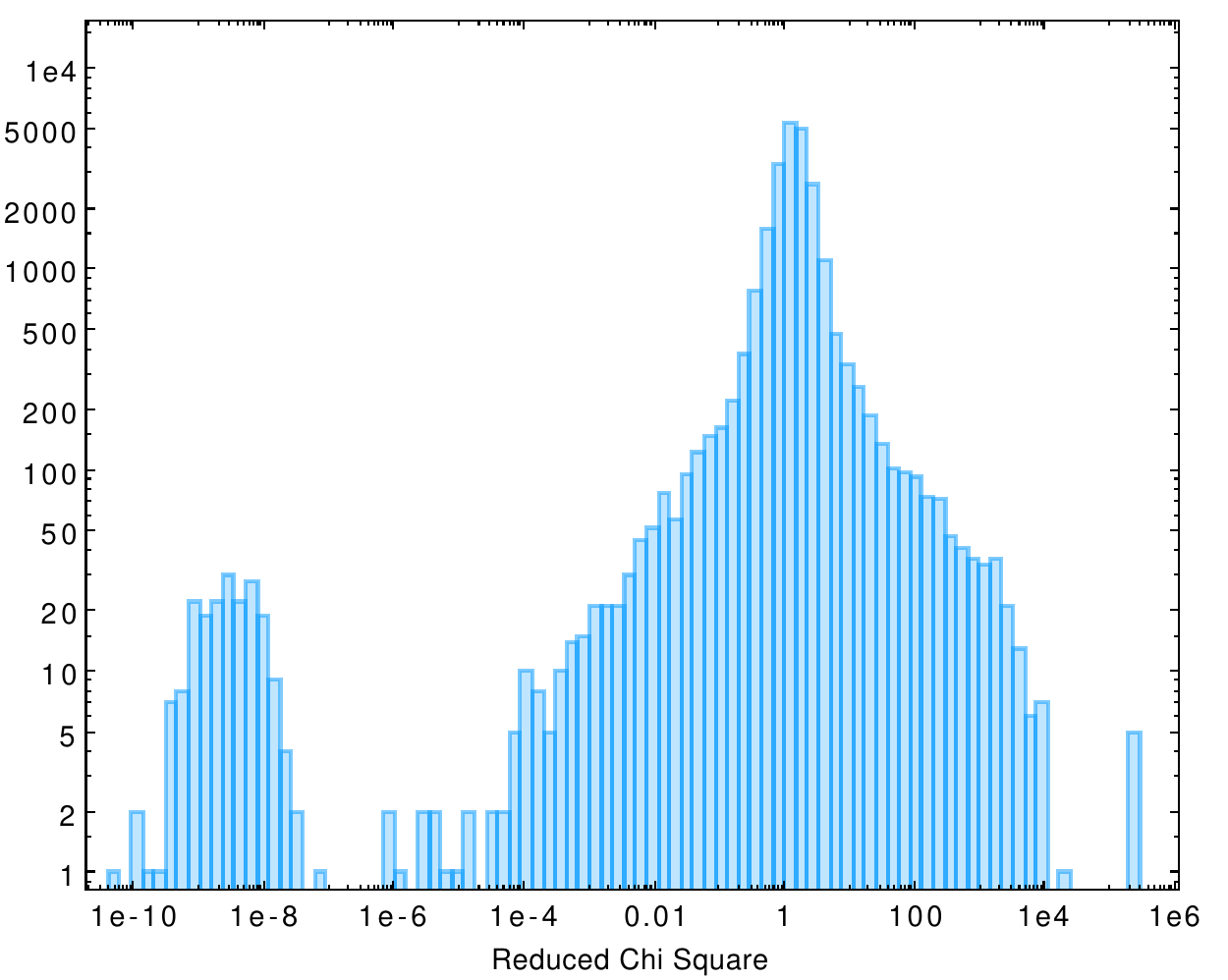}
    \end{minipage}
    \hfill
    \begin{minipage}{0.49\textwidth}
        \centering
        \includegraphics[trim={0 0 0 0},clip,width=\textwidth]{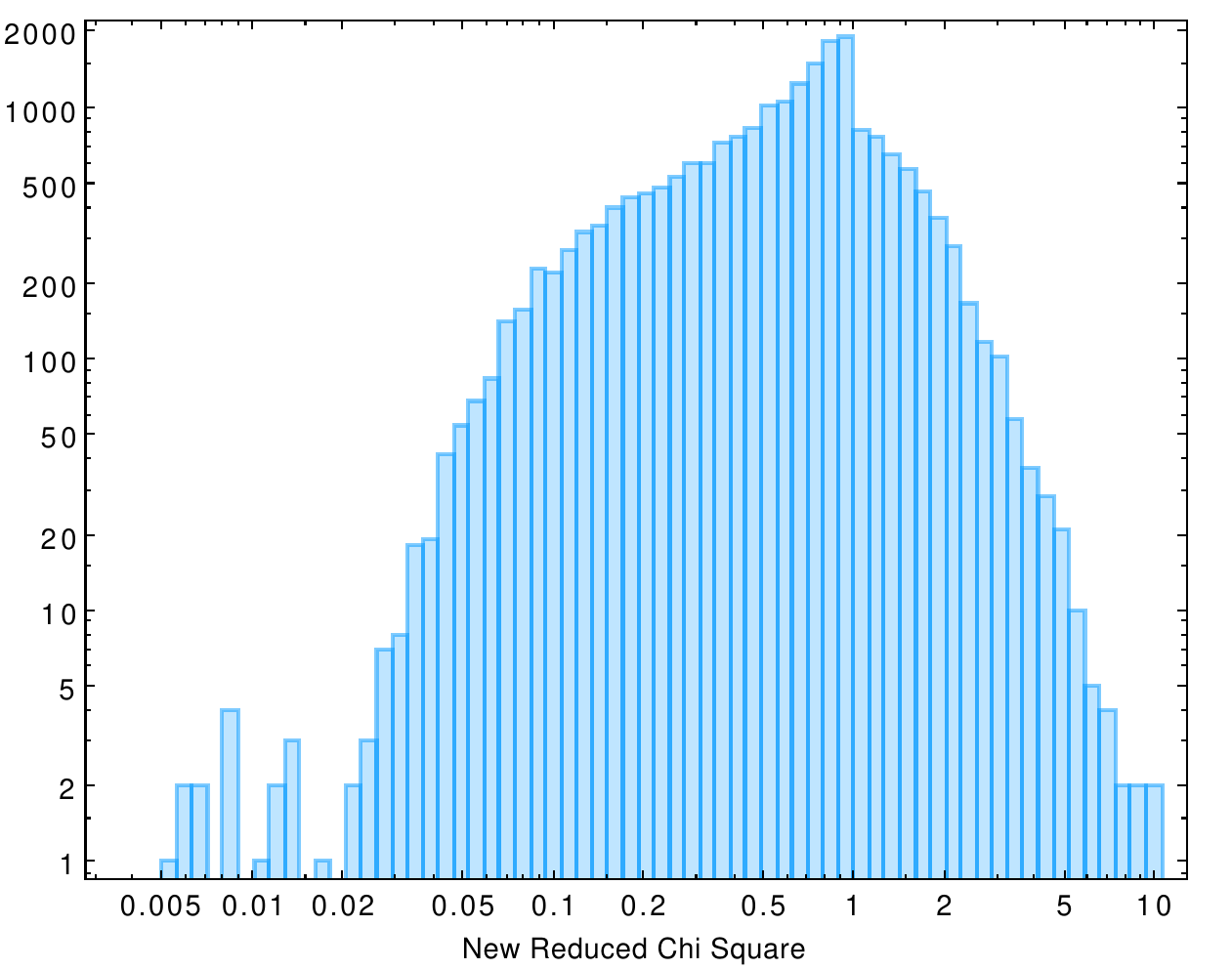}
    \end{minipage}
    
   \caption{{\bf Left}: Initial distribution of reduced chi-squared ($\chi^2_{\nu}$) values for spectral index fits, computed using flux density uncertainties directly extracted from the catalogue. {\bf Right}: Updated distribution of reduced chi-squared ($\chi^2_{\nu}$) values for the same sources, recalculated after applying uniform flux density uncertainties: $\pm5\%$ for sources with $S_{\nu} > 1$~mJy and $\pm10\%$ for sources with $S_{\nu} < 1$~mJy. Both distributions are shown on log--log scales to highlight the improvement in the fit quality after addressing underestimated uncertainties.}
    
    \label{fig:Figure 9}
\end{figure*}

\begin{figure*}
    \centering
    \begin{minipage}{0.49\textwidth}
        \centering
        \includegraphics[trim={5 0 0 0},clip,width=\textwidth]{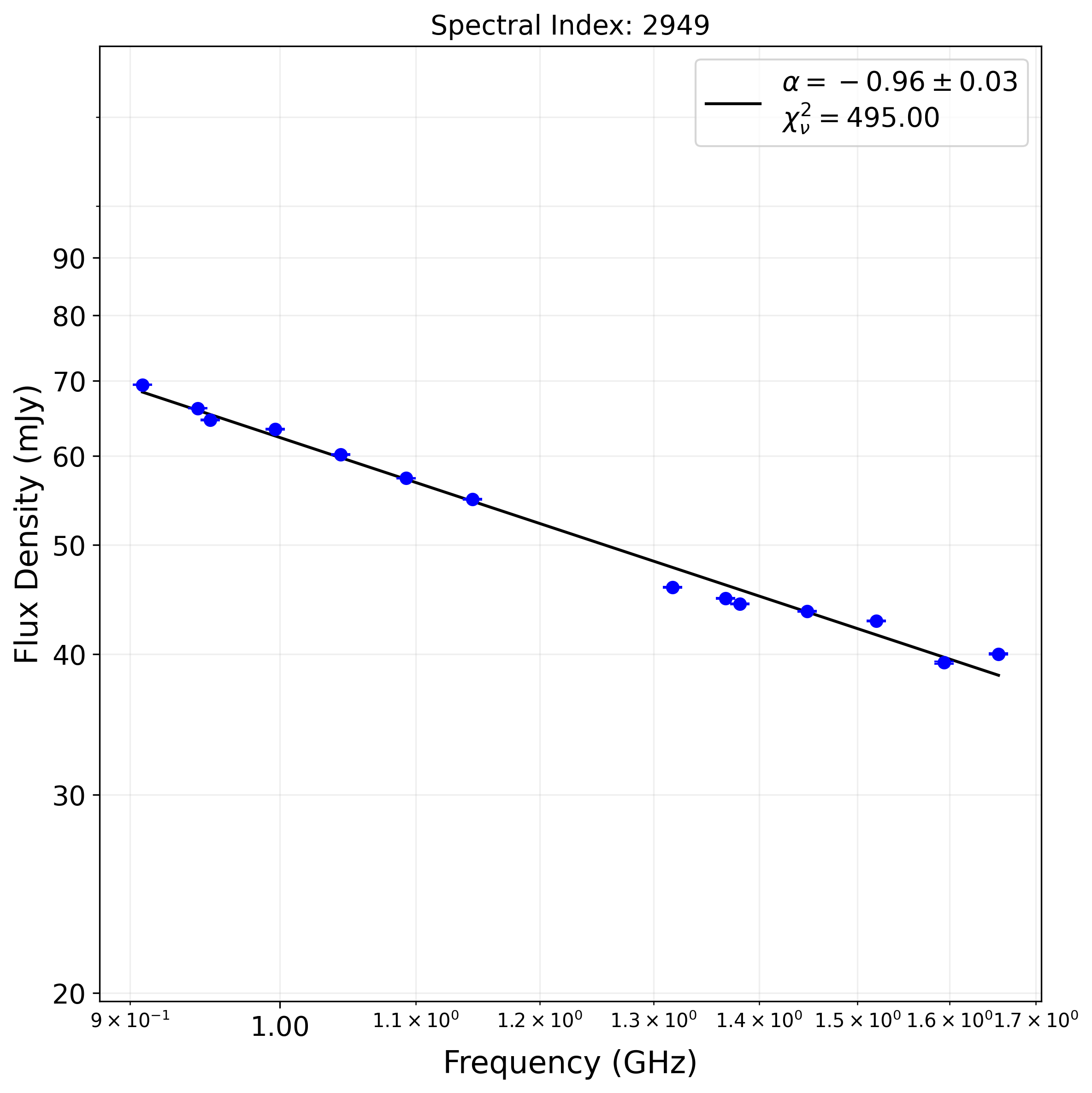}
    \end{minipage}
    \hfill
    \begin{minipage}{0.49\textwidth}
        \centering
        \includegraphics[trim={0 0 0 0},clip,width=\textwidth]{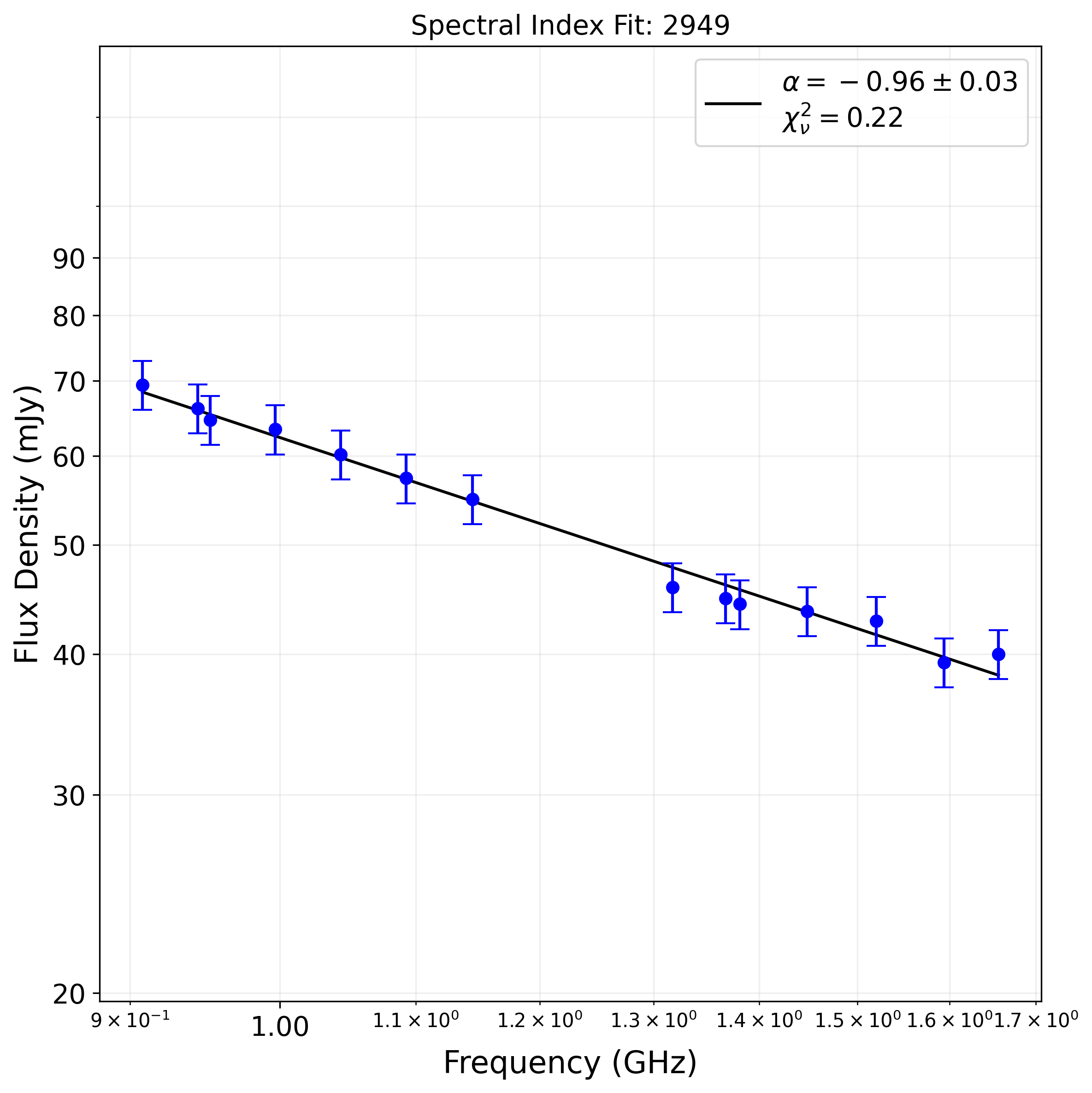}
    \end{minipage}
    
  \caption{{\bf Left}: Spectral index distribution of the source with \texttt{mainID 2949} using real flux density errors from the catalogue, with reduced chi-squared $\chi^2_{\nu} = 495.00$. 
    {\bf Right}: Same source, but assuming a flat $\pm5\%$ flux density error for all points, with $\chi^2_{\nu} = 0.22$.
    In both panels, flux density is plotted as a function of frequency on a log--log scale. The solid black line represents the orthogonal distance regression (ODR) fit in log space, from which the spectral index ($\alpha$) and reduced chi-squared ($\chi^2_{\nu}$) were derived. Vertical error bars indicate the assumed flux density uncertainties. All 14 available frequency measurements were used in both fits.}

    \label{fig:Figure 10}
\end{figure*}



\section{Results}
\label{Section 4}

Following the catalogue generation methodology outlined in Section \ref{Section 3}, the original 944\,MHz catalogue of 97,877 sources was filtered to 36,571 reliable sources distributed across the field. Most excluded sources were located at the field edges or in regions with \ac{BCE} sources, as shown in Figure~\ref{fig:Figure 3}.

The 1367\,MHz catalogue was generated using the same methodology as the 944\,MHz catalogue, from an initial 123,435 detected sources; this process yielded a filtered catalogue of 15,227 reliable sources.

\subsection{Positional Offsets}
\label{Results-PO}
Accurate positional measurements are critical for reliable source identification and multi-wavelength cross-matching in radio astronomy \citep{1998PASA...15..280T}. By comparing the positions of sources in our \ac{ASKAP} catalogues with those in the MeerKAT 1283\,MHz catalogue \citep{2024MNRAS.529.2443C} and the Milliquas catalogue \citep{2023OJAp....6E..49F}, we assess the astrometric precision of our data. Small positional offsets ensure that our catalogues can be effectively used for astrophysical studies, improving upon the positional accuracy of earlier \ac{ASKAP} surveys \citep{2019MNRAS.490.1202J}.


The positional comparison of \ac{ASKAP} 944\,MHz vs. MeerKAT 1283\,MHz shows mean $\Delta$RA\,=\,$-0\farcs33$ (STDEV\,=\,0.77) and $\Delta$DEC\,=\,0\farcs23 (STDEV\,=\,0.72) (see Figure~\ref{fig:posdiff}).

The positional comparison of \ac{ASKAP} 1367\,MHz vs. MeerKAT 1283\,MHz shows mean $\Delta$RA\,=\,$-0\farcs27$ (STDEV\,=\,0.63) and $\Delta$DEC\,=\,0\farcs14 (STDEV\,=\,0.64) (see Figure~\ref{fig:posdiff}).

To further assess the positional accuracy of our \ac{ASKAP} catalogues, we compared them to the latest version of the Milliquas catalogue~\citep{2023OJAp....6E..49F}, using a  2\arcsec\ crossmatching radius, consistent with the \ac{ASKAP} comparison. This comparison included both the \ac{ASKAP} 944\,MHz and 1367\,MHz catalogues, as well as the MeerKAT 1283\,MHz catalogue, to determine which telescope provides better positional accuracy relative to Milliquas.

The positional comparison of \ac{ASKAP} 944\,MHz vs. Milliquas shows mean $\Delta$RA\,=\,--0\farcs37 (STDEV\,=\,0.67) and $\Delta$DEC\,=\,0\farcs24 (STDEV\,=\,0.7)(see Figure~\ref{fig:posdiff}).

The positional comparison of \ac{ASKAP} 1367\,MHz vs. Milliquas shows mean $\Delta$RA\,=\,--0\farcs28 (STDEV\,=\,0.6) and $\Delta$DEC\,=\,0\farcs35 (STDEV\,=\,0.6)(see Figure~\ref{fig:posdiff}).

\subsection{Flux Densities Comparison}
\label{Results-FC}

To validate the flux calibration of our catalogue, we compared integrated flux densities of \ac{ASKAP} 1367\,MHz with the MeerKAT 1283\,MHz catalogue \citep{2024MNRAS.529.2443C}. These two surveys have closely matched central frequencies, enabling direct and meaningful comparisons.

We matched sources between the two catalogues and examined their flux density correlations across multiple signal-to-noise (S/N) thresholds see Figure~\ref{fig:Figure 5}). A total of 27,480, 20,524, 9,300, 4,289, and 1,777 sources were identified with detection significances above 3$\sigma$, 5$\sigma$, 10$\sigma$, 20$\sigma$, and 50$\sigma$, respectively. The flux densities of these subsets were compared using a linear regression model, yielding consistent slopes: 1.037 at 3$\sigma$, 1.048 at 5$\sigma$, and 1.050 at 50$\sigma$, indicating strong agreement across different S/N regimes (see Figure~\ref{fig:Figure 6}).


\subsection{Spectral Index }
\label{Results-SI}


The spectral index ($\alpha$) is a key diagnostic parameter in radio astronomy. We use the definition $S_\nu \propto \nu^\alpha$, where $S_{\nu}$ is the flux density at frequency $\nu$. It plays a central role in classifying radio sources and identifying their dominant radiation processes~\citep{2013MNRAS.429.2080W}. 


For instance, synchrotron-dominated sources such as AGN jets and SNRs typically display steep spectral indices ($\alpha < -0.3$), while thermal bremsstrahlung from \HII\ regions tends to produce flatter or even positive spectral indices~\citep{2019MNRAS.490.1202J}. The details about spectral index calculations and fitting methods are shown in the \ref{Fitting Method}.

In this study, spectral indices are derived using at least three distinct frequencies across flux density measurements from \ac{ASKAP} (944\,MHz, 1367\,MHz) and MeerKAT’s L-band (14 sub-bands, with bands 8 and 9 typically flagged due to RFI) spanning 908 to 1656\,MHz, enabling a robust evaluation of the astrophysical consistency and reliability of our catalogues.

In total, we obtained spectral index estimates for 21,442 sources~(see Figure~\ref{fig:Figure 7}). Among these, 16,172 sources have $\alpha < -0.3$, while 2,895 have $\alpha > -0.3$. A subset of 2,726 sources, each with flux density measurements at all 14 frequencies, $S_\nu > 1$\,mJy, spectral index uncertainty $\Delta \alpha < 0.2$, and reduced chi-squared $\chi^2_\nu < 1$, was used for the detailed spectral index versus flux analysis shown in Figure~\ref{fig:Figure 7}, as these criteria ensure the robustness of the fitted spectral indices. An example of a good spectral index model fit is shown in Figure \ref{fig:Figure 8}.

\section{Discussion}
\label{Section 5}

Compared to the earlier ASKAP-EMU observations of the \ac{SMC} by \citet{2019MNRAS.490.1202J}, which reported RMS noise levels of 186\,$\mu$Jy beam$^{-1}$ at 960\,MHz and 165\,$\mu$Jy beam$^{-1}$ at 1320\,MHz, detecting 4,489 and 5,954 sources, respectively, our \ac{ASKAP} surveys achieved substantially lower RMS noise levels of 26\,$\mu$Jy beam$^{-1}$ at 944\,MHz (see Figure~\ref{fig:SMC_IMAGE_944 MHZ}) and 28\,$\mu$Jy beam$^{-1}$ at 1367\,MHz (see Figure~\ref{fig:SMC image 1367}), with significantly higher source counts of 36,571 and 15,227, respectively. These improvements highlight the enhanced imaging sensitivity and refined calibration achieved in our observations of the \ac{SMC} field.

The statistical analysis of positional offsets reveals minor systematic differences between the \ac{ASKAP} and MeerKAT catalogues, with mean positional deviations remaining well within sub-arcsecond precision, indicative of good astrometric alignment. In contrast, \citet{2024MNRAS.529.2443C} reported $\sim$1\arcsec\ offsets when comparing MeerKAT positions to those from the earlier ASKAP-EMU survey \citep{2019MNRAS.490.1202J}, highlighting a clear improvement in astrometric accuracy in our current dataset. Furthermore, the comparison between \ac{ASKAP} and Milliquas source positions exhibits similar consistency, supporting a typical positional uncertainty of $\sim$0\farcs3 in our \ac{ASKAP} point source catalogue.

Our flux density comparison between \ac{ASKAP} 1367\,MHz and MeerKAT 1283\,MHz confirms a consistent linear relationship across varying S/N regimes. The fitted slope remains stable at $\sim$1.05, indicating a small but consistent offset in flux scale. Marginal histograms confirm a narrowing of flux distribution at higher confidence levels, consistent with reduced noise contamination.  While the results show excellent agreement for bright sources, a noticeable increase in scatter emerges at lower flux densities ($S_\nu < 1.2$\,mJy), particularly at the 3$\sigma$ and 5$\sigma$ levels (see Figure~\ref{fig:Figure 6}). This divergence is likely driven by differences in beam sizes, calibration uncertainties and underestimated uncertainties in faint sources. 

Compared to the earlier ASKAP–MeerKAT flux comparison in \citet{2024MNRAS.529.2443C}, which found agreement only at high S/N ($>$50$\sigma$), our analysis reveals a tighter correlation and broader consistency across lower thresholds. This improvement reflects both deeper imaging and refined calibration in our new \ac{ASKAP} dataset.

The results of our spectral index analysis show that the majority of sources in our catalogue exhibit spectral indices within the range $-1.7 < \alpha < +1.5$, in line with expectations from earlier radio continuum studies of the \ac{SMC}. Previous analyses, such as \citet{2024MNRAS.529.2443C}, reported typical spectral index distributions falling within $-2.5 < \alpha < +1.5$. The distribution peaks around $\alpha \approx -0.8$, characteristic of synchrotron-dominated emission from SNRs and AGNs, while a substantial population with $\alpha > -0.3$ likely corresponds to thermal emission from \HII\ regions and planetary nebulae.

To ensure the reliability of the spectral index measurements, we applied strict selection criteria, including detections at 14 frequency-flux measurements, uncertainties $\Delta \alpha < 0.2$, and reduced chi-squared values $\chi^2_{\nu} < 1$ (see Section~\ref{Reduced Chi Squared}). This filtering yielded 2,726 robust spectral index estimates across the \ac{SMC} field.

A direct comparison with the MeerKAT results of \citet{2024MNRAS.529.2443C} reveals excellent consistency: our mean spectral index is $\alpha = -0.62 \pm 0.35$, closely matching their reported $\alpha = -0.65 \pm 0.37$. This alignment confirms the accuracy of our \ac{ASKAP} flux density measurements and underscores the coherence between the \ac{ASKAP} and MeerKAT datasets.

Overall, the consistency of our spectral index measurements with previous studies highlights the reliability of our dataset and enhances the understanding of the radio-emitting population in the \ac{SMC}. These results provide a strong foundation for future multi-wavelength studies of the \ac{SMC}’s radio sources.

\begin{figure*}
    \centering
    \includegraphics[width=1\linewidth]{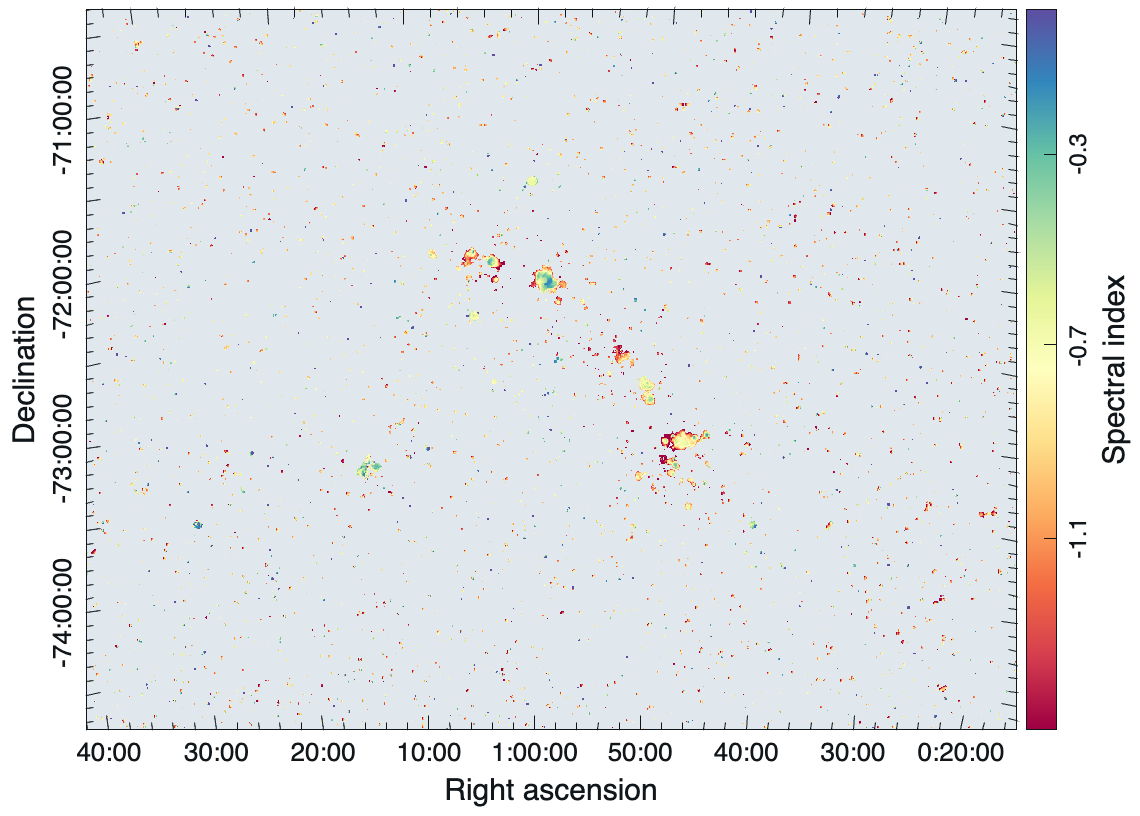}
    \caption{Spectral index image of the \ac{SMC} region using \ac{ASKAP} (944\,MHz and 1346\,MHz) and MeerKAT (1283\,MHz) images. All images have been convolved to the beam size of 20\arcsec$\times$20\arcsec\ and regridded to the largest pixel size of 2\arcsec$\times$2\arcsec. The colour bar is linearly scaled.}
    \label{fig:Figure 11}
\end{figure*}


\subsection{Spectral index mapping for merged images}
\label{sec:SI_map}

To complement our point source spectral index catalogue and provide spatial context for extended and diffuse emission, we constructed a spectral index map of the \ac{SMC} region using \textsc{miriad}\footnote{\url{http://www.atnf.csiro.au/computing/software/miriad/}}~\citep{1995ASPC...77..433S}. This map was derived from the \ac{ASKAP} 944\,MHz and 1367\,MHz images and the MeerKAT 1283\,MHz image (Figure~\ref{fig:Figure 11}), allowing pixel-wise characterization of spectral behaviour across the field.

To ensure uniform resolution and astrometric consistency, all input images were convolved to a common beam size of $20\arcsec\times20$\arcsec\ using \textsc{convol}, and regridded to a common pixel scale of $2\arcsec\times2$\arcsec\ using \textsc{regrid}. The \textsc{maths} task was then used to compute the spectral index values pixel by pixel. Although no formal uncertainty map was computed, the image was inspected to ensure that spectral indices were only evaluated in regions with significant emission above the local noise threshold at all three frequencies.

The spectral index map shows that the bulk of the emission throughout the \ac{SMC} displays steep spectral indices, largely in regions of significant star-forming complexes and \acp{SNR}. While full physical interpretation is outside the scope of this paper, the spectral index map marks a valuable data image. It offers a resolved tool for future studies to categorise extended sources, distinguish thermal and non-thermal emission, and explore environmental variations in the radio spectral properties of the \ac{SMC}.



\section{Conclusion}
\label{Section 6}

We summarise our results as follows:

\begin{itemize}
    \item \textbf{Point source identification}: 
    The new \ac{ASKAP} surveys of the \ac{SMC} field at 944 and 1367~MHz have identified a combined total of 36,571 and 15,227 point radio sources (Table~\ref{tab:ASKAP_944MHz_1367MHz_catalogue}).

   \item \textbf{Astrometric precision}: 
   Positional offsets between the \ac{ASKAP} and MeerKAT catalogues are within sub-arcsecond precision, with typical offsets of $\sim$0\farcs3, indicating improved astrometric alignment relative to earlier \ac{ASKAP} catalogue~\citep{2019MNRAS.490.1202J}.
    
    \item \textbf{Flux density measurements}: 
   Flux density measurements show strong agreement between \ac{ASKAP} 1367\,MHz and MeerKAT 1283\,MHz, with stable regression slopes near unity across a wide range of signal-to-noise thresholds. This confirms the consistency of flux calibration across instruments and detection regimes.

   \item \textbf{Spectral index distribution}: 
   The spectral index distribution of sources is consistent with previous studies~\citep{2024MNRAS.529.2443C}, with a mean of $\alpha = -0.62 \pm 0.35$. The distribution peak at $\alpha \approx -0.8$ reflects a predominance of non-thermal emission sources.
\end{itemize}

Overall, these catalogues aid in advancing our understanding of the \ac{SMC} population. These catalogues can be used as a foundation for future studies of the \ac{SMC} and enable a better understanding of the nature of the individual sources.


\section*{Acknowledgements}

This scientific work uses data obtained from Inyarrimanha Ilgari Bundara / the Murchison Radio-astronomy Observatory. We acknowledge the Wajarri Yamaji People as the Traditional Owners and native title holders of the Observatory site. CSIRO’s \ac{ASKAP} radio telescope is part of the Australia Telescope National Facility \footnote{\url{(https://ror.org/05qajvd42)}}. Operation of \ac{ASKAP} is funded by the Australian Government with support from the National Collaborative Research Infrastructure Strategy. \ac{ASKAP} uses the resources of the Pawsey Supercomputing Research Centre. Establishment of ASKAP, Inyarrimanha Ilgari Bundara, the CSIRO Murchison Radio-astronomy Observatory and the Pawsey Supercomputing Research Centre are initiatives of the Australian Government, with support from the Government of Western Australia and the Science and Industry Endowment Fund.
This research made use of the cross-match service provided by CDS, Strasbourg and of the SIMBAD database, operated at CDS, Strasbourg, France.



\section*{Data Availability}

The catalogues underlying this article are available on the CDS/VIZIER
(https://cds.u-strasbg.fr/) website.


\clearpage
\appendix
\section{Appendix}
\subsection{Fitting Method: Orthogonal Distance Regression (ODR)}
\label{Fitting Method}

To determine the spectral index, we performed a power-law fit using Orthogonal Distance Regression (ODR), which accounts for uncertainties in both flux density and frequency measurements. ODR is preferred over ordinary least squares (OLS) regression as it minimises perpendicular residuals rather than vertical ones, making it well-suited for cases where both variables have associated errors \citep{Boggs1990}.

The spectral index was derived by fitting a power-law function:

\begin{equation}
S_{\nu} = B_0 \nu^{\alpha}
\end{equation}

\noindent where $B_0$ is the normalisation factor (flux density at a reference frequency),
and $\alpha$ is the spectral index.

To linearise the fitting process, we rewrite this equation in logarithmic form:

\begin{equation}
\log S_{\nu} = \log B_0 + \alpha \log \nu
\end{equation}

This transformation allows for a linear regression approach while maintaining the power-law dependence.

We chose to use a simple power-law model (linear in log-log space) rather than higher-order polynomial fits. One of the primary reasons for this choice is to maintain consistency with MeerKAT surveys, which use the same power-law fitting method for spectral index estimation. Using the same approach ensures a direct and fair comparison between \ac{ASKAP} and MeerKAT data without introducing methodological biases. Additionally, given the typical spectral behaviour of synchrotron-emitting sources, a single power-law is sufficient for most cases and avoids overfitting the data.


\subsection{Implementation of the ODR Fit}
Flux densities and uncertainties were extracted from the catalogue for each available frequency, and an ODR power-law model was applied using the Python library \textsc{Scipy}\footnote{\href{https://docs.scipy.org/doc/scipy/reference/odr.html}{Scipy ODR Documentation}}~ \citep{2020SciPy-NMeth}.

The fitting process was implemented as follows:

\begin{enumerate}
    \item A power-law model was defined:
    \begin{equation}
    f(\nu) = B_0 \nu^{\alpha}
    \end{equation}
    where \( f(\nu) \) represents the modelled flux density as a function of frequency \( \nu \), \( B_0 \) is the normalisation constant corresponding to the flux density at a reference frequency, and \( \alpha \) is the spectral index. This function serves as the mathematical model that approximates the observed flux densities \( S_{\nu} \).
    
    \item ODR fitting was performed using scipy.odr.ODR:
    \begin{equation}
    \alpha = \text{ODR fit slope}, \quad B_0 = \text{ODR fit intercept}
    \end{equation}
    
    \item The best-fit parameters ($\alpha, B_0$) and uncertainties were extracted.
\end{enumerate}

\subsection{Assessment of Fit Quality}
\label{Reduced Chi Squared}
To ensure reliability, we evaluated the fit quality using the reduced chi-squared statistic:

\begin{equation}
\chi^2_{\nu} = \frac{1}{N - p} \sum \frac{(S_{\nu, \text{obs}} - S_{\nu, \text{model}})^2}{\sigma^2}
\end{equation}

\noindent where, $N$ is the number of flux density measurements for a given source, $p$ is the number of free parameters in the model (here, $p = 2$ for $\alpha$ and $B_0$), $S_{\nu, \text{obs}}$ is the observed flux density, $S_{\nu, \text{model}}$ is the flux density predicted by the power-law fit and  $\sigma$ is the flux density uncertainty\citep{Bevington1969}.

The distribution of reduced chi-squared values ($\chi^2_{\nu}$) for all fitted sources is shown in Figure~\ref{fig:Figure 9}.

To exclude poorly constrained fits, we applied the following selection criteria:

\begin{itemize}
    \item Only sources with flux density measurements at all 14 available frequencies were included.
    \item Only sources with flux density greater than 1 mJy were considered.
    \item Spectral indices with uncertainty $\Delta \alpha < 0.2$ were considered reliable.
    \item Sources with reduced chi-squared $\chi^2_{\nu} < 1$ were selected to ensure a good fit.
\end{itemize}

The final selection of sources after applying these quality filters is visualised in the left panel of Figure~\ref{fig:Figure 8}.
This scatter plot shows the distribution of spectral indices as a function of integrated flux density. The data points are colour-coded based on their spectral index uncertainty ($\Delta \alpha$), illustrating the level of fit reliability across different flux levels. The selection criteria ensure that the retained sources have well-constrained spectral indices and statistically significant fits.




\newcommand\eprint{in press }

\bibsep=0pt

\bibliographystyle{aa_url_saj}

{\small

\bibliography{Final_Manuscript}
}

\clearpage

{\ }

\clearpage

{\ }

\newpage

\begin{strip}

{\ }




\vskip-20mm

\naslov{Nova ASKAP radio-kontinuum istra\zz iva\nj a Malog Magelanovog Oblaka (MMO).}


\authors{
O. K. Khattab$^1$,
{\rrm M. D. Filipovi\cc}$^1$,
Z. J. Smeaton$^1$,
R. Z. E. Alsaberi$^{2,1}$,
E. J. Crawford$^{1}$,
}
\authors{
D. Leahy$^3$,
S. Dai$^{4,1}$
and N. Rajabpour$^1$
}


\vskip3mm


\address{$^1$Western Sydney University, Locked Bag 1797, Penrith South DC, NSW 2751, Australia}
\address{$^2$Faculty of Engineering, Gifu University, 1-1 Yanagido, Gifu 501-1193, Japan}
\address{$^3$Department of Physics and Astronomy, University of Calgary, Calgary, Alberta, T2N IN4, Canada}
\address{$^{4}$Australian Telescope National Facility, CSIRO, Space and Astronomy, P.O. Box 76, Epping, NSW 1710, Australia}

\Email{22197951@student.westernsydney.edu.au}

\vskip3mm


\centerline{{\rrm UDK} \udc}


\vskip1mm

\centerline{\rit Uredjivaqki prilog}

\vskip.7cm

\baselineskip=3.8truemm

\begin{multicols}{2}

{
\rrm

\abstract{
Predstavljamo dva nova radio-kontinuum pregleda u okviru \textsc{ASKAP POSSUM} istra{\zz}ivanja u pravcu Malog Magelanovog Oblaka (MMO). 
Nove dve liste radio izvora koje su proizvedene iz ovih pregleda sadr{\zz}e 36,571 radio-kontinuum izvora posmatranih na 944\,MHz i 15,227 izvora na 1367\,MHz, sa veliqinama snopa od 14\farcs5$\times$12\farcs2 i 8\farcs7$\times$8\farcs2. 
Za kreira{nj}e ovih kataloga taqkastih izvora koristili smo softver \textsc{Aegean}, a zajedno sa pretnodno objavljenim \textsc{MeerKAT} katalogom taqkastih izvora, procenili smo spektralne indekse za celu populaciju zajedniqkih radio taqkastih izvora. 
Uporedjivanjem na{\ss}ih \textsc{ASKAP} kataloga sa \textsc{MeerKAT} katalogom, na{\ss}li smo 21,442 i 12,654 taqkastih radio izvora zajedniqkih za preglede na 944\,MHz i 1367\,MHz a u okviru 2\arcsec\ tolerancije.
Ovaj katalog taqkastih radio izvora sa radio teleskopa nove generacije kao {\ss}to su \textsc{ASKAP}, pomo{\cc}i{\cc}e da bolje izuqimo razliqite populacije radio izvora.
%

}}
\end{multicols} 

\end{strip}

\end{document}